%% file: article.tex
\documentclass[12pt]{article}

\usepackage{libertinus} 

\usepackage{amsmath}    
\usepackage{amssymb}    
\usepackage{amsthm}     
\newtheorem{proposition}{Proposition}
\newtheorem*{proposition*}{Proposition}
\newtheorem*{remark}{Remark}

\DeclareMathOperator*{\argmin}{arg\,min}

\usepackage[style=authoryear]{biblatex}
\usepackage[left=1.5cm,right=1.5cm,top=2.5cm,bottom=2.5cm]{geometry}
\usepackage{placeins}

\usepackage{fancyhdr}
\usepackage{tocloft}
\usepackage{graphicx}   
\usepackage{subcaption} 
\usepackage{tikz}
\usetikzlibrary{matrix, patterns} 

\usepackage{tabularx}   
\usepackage{booktabs}   
\usepackage{makecell}

\usepackage{algorithm}
\usepackage{algorithmic}

\usepackage[textwidth=4cm, textsize=footnotesize]{todonotes}
\usepackage{setspace}   
\usepackage{soul,color} 
\usepackage{cancel}     

\usepackage{tasks}      
\usepackage[colorlinks=true, linkcolor=blue, citecolor=blue, urlcolor=blue]{hyperref}



\begin{document}

\begin{center}
{\Large
	{\sc Optimal Transport-Based Clustering of Attributed Graphs with an Application to Road Traffic Data}
}
\bigskip

Ioana Gavra $^{1}$ \&  Ketsia Guichard-Sustowski  $^{2}$ \& Loïc Le Marrec $^{3}$
\bigskip

{\it
$^{1}$ Univ Rennes 2, IRMAR - UMR CNRS 6625, F-35000 Rennes, France, ioana.gavra@univ-rennes2.fr \\
$^{2}$ Univ Rennes, IRMAR - UMR CNRS 6625, F-35000 Rennes, France, ketsia.guichard@univ-rennes.fr\\
$^{3}$ Univ Rennes, IRMAR - UMR CNRS 6625, F-35000 Rennes, France, loic.lemarrec@univ-rennes.fr
}
\end{center}
\bigskip


{\bf Abstract.} In many real-world contexts, such as social or transport networks, data exhibit both structural connectivity and node-level attributes. For example, roads in a transport network can be characterized not only by their connectivity but also by traffic flow or speed profiles. Understanding such systems therefore requires jointly analyzing the network structure and node attributes, a challenge addressed by attributed graph partitioning, which clusters nodes according on both connectivity and attributes. In this work, we adapt transport-based approaches based on Gromov--Wasserstein (GW) discrepancy. We investigate how GW methods, traditionally used for general-purpose tasks such as graph matching, can be specifically adapted for node partitioning, an area that has been relatively underexplored.  In the context of node-attributed graphs, we introduce an adaptation of the Fused GW method, offering theoretical guarantees and the ability to handle heterogeneous attribute types. Additionally, we propose to incorporate distance-based embeddings to enhance performance.  The proposed approaches are systematically evaluated using a dedicated simulation framework and illustrated on a real-world transportation dataset. Experiments investigate the influence of target choice, assess robustness to noise, and provide practical guidance for attributed graph clustering. In the context of road networks, our results demonstrate that these methods can effectively leverage both structural and attribute information to reveal meaningful clusters, offering insights for improved network understanding.

{\bf Keywords}: attributed graph, graph partitioning, optimal transport, Gromov-Wasserstein, road network partitioning.

\tableofcontents 

\input{sections/1_introduction}

\input{sections/2_methods}
\input{sections/3_simulations}
\input{sections/4_applications}

\section{Discussion}

In this work we have explored several distance-based methods for clustering attributed graphs. We illustrate how these methods can effectively leverage both structural and attribute information to reveal meaningful clusters in the context of road networks partitioning, offering insights for improved network understanding. Even though in the present work we focused on traffic related data, one of the advantages of the proposed algorithms is their inherent ability to deal with a large class of attributes lying in general metric spaces.
After a thourough description of each method and of the links between them, we assess their performances on various simulated scenarios.
In our numerical experiments, the methods based on the Gromov--Wasserstein divergence seem to perform better than other classical clustering methods like $k$-means or spectral clustering. 
We propose different strategies for defining the target graph for the OT based methods and evaluate their impact on the performances. The equidistant graph seems like a good default choice and setting the distance between the vertices of the target as the average distance in the source graph often works better than setting it to the maximal value. 
While some practical guidelines that can help choosing the target graph are illustrated by our numerical experiments, conducting a theoretical study to better understand the impact of the target on the clustering performance could be an interesting perspective. 

Another practical possible extension of the present work, can be the optimization of the target structure, for example by using the strategy described by \textcite{peyre2016gromov} for the computation of  Gromov--Wasserstein barycenters. 

We also propose using a feature map representation of the data within the OT based methods and illustrate via different simulation settings when this strategy enhances the performance of the algorithms. This type of strategy seems to be a promising lead, that would deserve a deeper exploration in a future work. One straight forward extension in this direction is to implement the use of an adaptive embedding in the spirit of \textcite{li2017locally} where the parameter $\alpha$ (used to combine the structural and attribute-related information) is optimized.

A limitation inherent to the proposed methods is the size of the graphs on which they can be applied.  While the computational cost is not an issue for the traffic road applications considered in this paper, it might become one in application fields that need to handle huge networks (for example graphs with millions of nodes). This limitation is quite common for OT based methods.  A possible approach to improve upon it, could be the exploration and adaptation of recently developed alternative transport plans or transport based divergences that are less computationally expensive, like Differentiable Generalized Sliced Wasserstein Plans \parencite{chapel2025differentiable} or Sliced Gromov--Wasserstein \parencite{titouan2019sliced}.

\paragraph*{Acknowledgement}The authors gratefully acknowledge V\'{e}ronique Thelen for the insightful discussions and continuous support, and the association Agis Ta Terre for their keen interest and active involvement in the field.

\paragraph*{Fundings}This work was conducted within the France 2030 program, Centre Henri Lebesgue ANR-11-LABX-0020-01. This project was supported by the participatory research platform of TISSAGE - Science avec et pour la société.

\FloatBarrier
\input{sections/X_appendix}

\FloatBarrier
\printbibliography
\end{document}

%% file: sections/1_introduction.tex
\section{Introduction}

\subsection{Context} 

Graph clustering aims to better understand large networks and discover communities within them. Furthermore, it might help identify vertices with a central position or specific edges between different clusters. This problem has broad applications across various fields, including social sciences, biology, and medicine \parencite{fortunato2010community}. For instance, in biology, clustering protein-protein interaction networks helps to better understand the organization and functional processes of cells \parencite{bhowmick2015clustering}. 
In social network analysis, clustering has been extensively studied to detect cohesive communities of users \parencite{bothorel2015clustering}.

Transport is another domain where graph clustering could address a common challenge: road network partitioning. Managing complex and extensive road networks may be facilitated by dividing them into smaller subnetworks, enabling the implementation of optimized control strategies. As outlined by \textcite{ji2012spatial}, the main objectives of such partitioning are to minimize the variance of link densities within each subnetwork 
and to reduce the total number of clusters, thereby enhancing network interpretability and simplifying control design. Several methodologies have been proposed: some focus solely on either the topological aspect of the road network or the traffic attributes \parencite{lin2020road}, while others consider both simultaneously, primarily through spectral clustering and multi-step approaches \parencite{anwar2014spatial, saeedmanesh2016clustering}.

Node clustering in a graph involves grouping vertices into distinct clusters. For instance, when considering the graph's topology, clusters could be formed by identifying groups with a higher density of connections within the cluster compared to between clusters. However, topology alone may not be sufficient to uncover meaningful communities. Without metadata, structural information might highlight groups that differ from those identified based on similar attributes \parencite{hric2014community}. Attributed graph clustering is at the confluence of graph clustering and multidimensional clustering of tabular data: the goal is to partition the graph into several clusters with a cohesive intra-cluster structure and homogeneous attribute value. The problem is particularly challenging because structural similarities and attribute similarities are often seen as independent or even conflicting objectives.

The nature of the attributes may vary a lot from one application domain to another. 
For example, in social network analysis, node attributes are typically discrete variables (such as age) or descriptive labels (e.g., gender or interests). However, in road network partitioning, attributes often take on more specific forms. Beyond length or qualitative characteristics, roads could be described by traffic flow metrics, such as average traffic flow (vehicles per time unit), average density (vehicles per length unit), or speed profiles. These traffic features are frequently represented as time series, functional data, or histograms.

To encompass a large diversity of applications domains, our objective is to adapt and study partitioning methods for attributed graphs, with attributes lying in a general metric space. In this context, particular attention is given to distance-based methods, especially within the framework of optimal transport-based graph partitioning. Simulated experiments enable a systematic comparison of the proposed methodologies under various conditions, including sensitivity to graph topology. They are compared with other graph partitioning approaches, namely Fr\'{e}chet $k$-means, adapted to the graph setting, and spectral clustering, before being applied to real-world data.

Our main contributions are summarized as follows:
\begin{itemize}
\item We propose new adaptations of Gromov--Wasserstein (GW) methods, originally designed for general-purpose graph comparison, to the task of graph partitioning using distance matrices;
\item Optimal transport-based partitioning methods project the initial source graph on a certain (usually small) target graph. We propose different strategies for choosing this target and study its impact  on the performance of the methods through numerical experiments;
\item In particular, we focus on node-attributed graphs by introducing an adaptation of Fused Gromov--Wasserstein (FGW) method for node partitioning, supported by theoretical guarantees. This method furnishes a simultaneous fusion version, and thus allows a comparison with the \textit{a priori} GW approach. The proposed methods are generic and could handle arbitrary attribute types or combinations of attributes with a defined distance;
\item We propose new versions of these methods, using a distance-based embedding of the data and we illustrate how this could enhance their performance;
\item We compare the proposed methods, using both a dedicated simulation framework and a real-world dataset from transport and mobility applications. The best-performing GW-based method demonstrate improved performance in comparison to standard methods such as spectral clustering in our simulation framework.
\end{itemize}

\subsection{Problem Description}

An undirected node-attributed graph is considered as the triple $G = (V,E,A)$. $V$ is the vertex set, $E$ the set of edges, and $A$ the set of attributes associated to the vertices. It is assumed that $V=\{v_1,\ldots, v_N\}$ is endowed with  a probability measure $\boldsymbol{\mu} =(\mu(v_1), \dots, \mu(v_N))$. This probability distribution could, for example, model the importance of each node of the network. Furthermore, each edge $e_{ij}\in E$, connecting $v_i$ and $v_j$, is endowed with a length $l_{ij}>0$. 
Finally, the attributes belong to a metric space $(\mathcal{X},d_\text{A})$, with $d_\text{A}$ a well-defined distance function. They could be of various types, such as labels, have continuous or discrete values and $\mathcal{X}$ could be the Cartesian product of distinct subspaces, each equipped with an appropriate distance function.



For illustration, attributes of two kinds are used: functional and histogram-based. In this setting, the attribute space $\mathcal{X}$ splits into two subspaces: $\mathcal{X}=\mathcal{C} \times \mathcal{H}$, where $\mathcal{C}$ is a space of curves equipped with a distance $d_\text{C}$ and $\mathcal{H}$ is a space of histograms endowed with a distance $d_H$.

Although a soft clustering problem could be considered, the present work focuses on a hard clustering setting. With that in mind, in what follows, the clustering of an attributed graph aims to partition the set of nodes $V$ into $k$ subsets $C_1, \dots, C_k$ of $G$ such that $\cup^k_{i = 1} C_i = V$ and $C_i \cap C_j = \emptyset$ for $i \neq j$.

\subsection{Related Work} 

Various strategies have been proposed specifically for clustering node-attributed graphs, and comprehensive surveys exist on this topic, both on node and edge partitioning \parencite{bothorel2015clustering} and on the closely related problem of community detection \parencite{chunaev2020community}. The latter classifies the methods into three main categories, summarized here.

\paragraph{Early-fusion Methods} Early-fusion methods combine structural and attribute information prior to the community detection process. Several strategies exist: some approaches incorporate attributes directly into the graph by translating them into edge weights that capture attribute similarities \parencite{neville2003clustering}, thereby reducing the problem to clustering a weighted graph. Conversely, other approaches discard the explicit network structure and instead encode structural information into pairwise node distances, which are then combined with attribute similarities \parencite{combe2012combining}. Finally, other methods rely on node-embedding techniques, for which a rich body of literature is available \parencite{cui2018survey}.

\paragraph{Simultaneous Fusion Methods} Simultaneous fusion methods combine structure and attribute during the process of community detection. Several strategies have been proposed. A first modifies the objective functions of well-known algorithms such as Louvain or Normalized Cut. For instance, \cite{boobalan2016graph} use a Local Outlier Factor approach to form clusters, which are then partitioned using attribute similarity. Another approach relies on matrix factorization, in particular non-negative matrix factorization (NNMF), which approximates high-rank non-negative matrices by lower-rank factors to uncover latent clusters \parencite{berahmand2022graph}. 

\paragraph{Late Fusion Methods} Late fusion methods apply community detection independently on the structural and attribute spaces, and then merge the resulting partitions into a single consensus \parencite{huang2016consensus}. This combination is typically achieved through consensus clustering algorithms, which reconcile multiple partitions into a unified solution.

Along this useful classification, this survey underlines the absence of a universally preferred method in the field, as well as the lack of consensus on the impact of fusing structure and attributes, particularly regarding when such fusion is truly beneficial. Therefore, our work explores both early- and simultaneous-fusion strategies.

In particular, our work seeks to both evaluate and advance graph clustering techniques grounded in optimal transport. Section~\ref{ot_graph} develops GW methods, which form the core building blocks of our algorithms, while existing extensions for graph partitioning are further detailed in Section~\ref{existing_gw_methods}. Another optimal-transport-based approach for graph partitioning exists \parencite{ni2019community} but, to the best of our knowledge, is not directly applicable to attributed graphs, though still worth mentioning. This latter, based on Ollivier-Ricci curvature, leverages optimal transport to assign curvature values to edges, reflecting how probability mass spreads locally between nodes. Positive curvature typically corresponds to well-connected intra-community links, whereas negative curvature highlights inter-community “bridges”. By identifying and removing negatively curved edges, the method aims to reveal the community structure in complex networks.

\subsection{Distance-based Methods}

This article focuses on distance-based methods, which may be implemented within early-fusion, simultaneous, or even late-fusion frameworks. The notion of distance (or similarity) is central to clustering \parencite{jain1999data}, as it provides a metric (or quasi-metric) on the feature space to quantify pattern similarity. Distance-based clustering methods are particularly appealing in this context because they could be applied to general metric spaces without requiring additional structure (e.g., Euclidean geometry).

From this perspective, constructing a distance on $V$ that integrates both structural and attribute similarities could be a natural first step for clustering attributed graphs. \textcite{combe2012combining} proposes encoding both types of information into a unified distance function between nodes. This function could then be used with distance-based clustering methods:
\begin{equation}
    d_{\alpha}(v_i, v_j) = \alpha \cdot d_{\text{S}}(v_i, v_j) + (1 - \alpha) \cdot d_{\text{A}}(v_i, v_j)
\label{d_def}
\end{equation}

where $d_{\text{S}}(v_i, v_j)$ denotes the normalized structural distance between $v_i$ and $v_j$, $d_{\text{A}}(v_i, v_j)$ represents the distance between their attributes, and $\alpha \in [0, 1]$ is a weighting parameter.

Note that the structural and attributes distances may be combined in various ways from the product space $(V,d_{\mathrm{S}}) \times (A,d_{\mathrm{A}})$, and not only through this linear combination. For instance, a valid candidate for $d_{\alpha}$ could be $\|(d_{\mathrm{S}}, d_{\mathrm{A}})\|_{2,\alpha} 
= \sqrt{\alpha \cdot d_{\mathrm{S}}^2(v_i, v_j) + (1 - \alpha) \cdot d_{\mathrm{A}}^2(v_i, v_j)}$. This article will consider the initial definition given in equation (\ref{d_def}).

Let $\mathbf{D}_{\alpha} = [d_{\alpha}(v_i, v_j)]_{1 \leq i,j \leq N}$ represents the symmetric distance matrix representing the pairwise distances between all nodes. In the same manner, we define $\mathbf{D}_S$ as the structural distance matrix and $\mathbf{D}_A$ as the attribute distance matrix.

%% file: sections/2_methods.tex
\FloatBarrier
\section{Methodology}

Constructing a distance between nodes thus provides a unified framework, enabling the application of both traditional partitioning methods (such as $k$-means) and optimal transport-based approaches specifically tailored to graphs, while also allowing to investigate the links between these methods. $k$-means is first used to illustrate distance-based approaches on graphs, followed by the introduction of optimal transport-based methods, before presenting the main contributions of this work.

\subsection{k-means for Graphs}\label{kmeans_section}

$k$-means \parencite{mcqueen1967some} is a well known method for clustering data. Its popularity comes from its intepretability and from the simplicity of the Lloyd's algorithm \parencite{lloyd1982least}, which is typically employed to find a locally optimal solution in vector spaces. However, applying $k$-means directly to graphs is challenging because computing means is not straightforward in a graph setting. To address this, an adaptation that relies on using Fr\'{e}chet means within the widely used Lloyd's algorithm is proposed. The Fr\'{e}chet mean was introduced as an extension of the expected value of a random variable to general metric spaces \parencite{frechet1948elements}.

\subsubsection{Graph Partitioning with k-Fr\'{e}chet Means}
Consider a compact metric space $(\Omega,d)$ endowed with a probability measure $\mu$ that has a finite second order moment. The associated Fr\'{e}chet-$k$ means functional $U_{\mu,k}:\Omega^k\to\mathbb{R}^+$ can be defined as:
\begin{equation*}
    U_{\mu,k}(m_1,\ldots,m_k)=\int_{\Omega}d(x,\{m_1,\ldots,m_k\})^2\mu(\mathrm{d}x),
\end{equation*}
where $d(x,\{m_1,\ldots,m_k\})$ denotes the usual point to set distance, $d(x,\{m_1,\ldots,m_k\})=\min_i d(x,m_i)$. Minimizing this function over $\Omega^k$ gives $k$ optimal centers for the probability measure $\mu$. These centers could be used to define a partition of  $\Omega$ by computing the associated Voronoi cells (each point of $\Omega$ is assigned to the closest center). Although this is out of scope for this paper, one could consider a more general problem, based on moments of order $q$, by defining $U_{\mu,k,q}(m_1,\ldots,m_k)=\int_{\Omega}d^q(x,\{m_1,\ldots,m_k\})\mu(\mathrm{d}x)$.   \\
Now consider $N$ data points $\{x_1, \dots, x_N\}$ in $\Omega$. A Fr\'{e}chet-$k$ means clustering problem aims to partition these points into $k$ clusters $\{C_1, \dots, C_k\}$ that minimize the total within-cluster dissimilarity:
\vspace{-8pt}
\begin{equation*}
    \min_{\{C_i\}_{i=1}^k, \{m_i\}_{i=1}^k} \sum_{i=1}^k \sum_{x \in C_i} d(x, m_i)^2 \mu(x)
\end{equation*}
where $C_i \subset \{x_1, \dots, x_N\}$ is the set of points assigned to the $i$-th cluster and $m_i$ is a Fr\'{e}chet mean of the cluster $C_i$, i.e., $$m_i \in \arg \min_{y \in \Omega} \sum_{x \in C_i} d(x,y)^2 \mu(x).$$
In the graph setting, one could consider $\Omega = V$, the set of vertices of the graph. More specifically, for an attributed graph, the distance $d_{\alpha}$ encodes both the graph structure and the discrepancies between the node attributes. This approach could then be used for node-attributed graph clustering by setting $(\Omega,d)=(V,d_{\alpha})$. A Fr\'{e}chet-$k$ means algorithm for this setting is described in Algorithm \ref{algo_k}. This algorithm is based on the Lloyd heuristic, but at each iteration, the centers of the clusters are updated using a Fr\'{e}chet mean instead of the classical weighted arithmetic mean that is not defined in our setting. 
When the data points are in an Euclidean space, $\Omega=\mathbb{R}^d$, and the distance $d$ is given by the $l_2$ norm, $d(x,y)=\|x-y\|_2$, the procedure described in Algorithm \ref{algo_k} reduces to the standard Lloyd $k$-means algorithm.
\begin{remark}
As in the standard Euclidean framework, the objective function $U_{\mu,k}$ decreases along the iterates $\textbf{m}_n=(m_1^n,\ldots,m_k^n)$ produced by the Algorithm \ref{algo_k} and thus the sequence $(U_{\mu,k}(\textbf{m}_n))_{n\ge 0}$ is convergent. 
\end{remark}
\paragraph{Sensitivity to Initialization} Note that this method requires $k$ initial centers $m_1, \dots, m_k$. It is well known that randomly selecting $k$ points from the data as the initial cluster medoids for the Lloyd algorithm may easily lead to an inefficient local optimum \parencite{pena1999empirical}. To improve the quality of the results, these initial medoids could be chosen using dispersive strategies, such as $k$-means++ \parencite{arthur2007k}. This initialization method is used for our numerical simulations in Section \ref{sec : Results}. 

\paragraph{Computational Complexity} In Algorithm \ref{algo_k}, all pairwise distances are precomputed and stored in the matrix $\mathbf{D}_{\alpha}$. At each iteration $n$, the assignment step requires $\mathcal{O}(Nk)$ operations, while the update step costs $\mathcal{O}(N^2)$ in the worst case, if a single cluster contains nearly all nodes, or $\mathcal{O}(N^2/k)$ for balanced clusters. Hence, the total time complexity is $\mathcal{O}(n(Nk + N^2))$ in the worst case and $\mathcal{O}(n(Nk + N^2/k))$ in the balanced case. 

\begin{algorithm}
    \small
    \caption{Fr\'{e}chet-$k$ means clustering for node-attributed graph clustering}
    \begin{algorithmic}
    \STATE \textbf{Input:} An attributed weighted graph $G = (V, E, A)$ characterized by its pair-wise distance matrix $\mathbf{D}_{\alpha} = [d_{\alpha}(v_i, v_j)]_{1 \leq i,j \leq N}$ and its probability distribution $\mu$, $k$ the number of desired subsets, initial clusters centers  $m_1, \dots, m_k$ and a maximum number of iterations
    \REPEAT
        \FOR{each node $v_i \in V$}
            \STATE Assign $v_i$ to the nearest cluster center, i.e., $\arg \min_{1 \leq j \leq k} d_{\alpha}(m_j, v_i)$
        \ENDFOR
        \FOR{each cluster $C_j$}
            \STATE Update the cluster center: $m_j = \arg \min_{x \in C_j} \sum_{v \in C_j} d_{\alpha}^2(v, x) \mu(v)$
        \ENDFOR
    \UNTIL{Convergence (no change in cluster centers) or maximum number of iterations is reached.}
    \STATE \textbf{return} Subsets $\{C_1, \dots, C_k\}$ and their centers $\{m_1, \dots, m_k\}$
    \end{algorithmic}
    \label{algo_k}
\end{algorithm}

\subsubsection{An Optimal Transport Perspective on
$k$-Means}

Before addressing optimal transport adaptations for graph partitioning, it is worth noting that $k$-means could be interpreted as a Wasserstein barycenter problem, as already highlighted by \textcite{cuturi2014fast} and \textcite{ho2017multilevel}.

Let $(\Omega, d)$ be a metric space. A Wasserstein barycenter (\cite{agueh2011barycenters}) of $M$ weighted probability measures $\{\mu_1, \dots, \mu_M\}$, supported in $\Omega$, is the probability distribution $\nu$ that minimizes:
\begin{equation*}
    \argmin_\nu \sum_{i = 1}^M W^2_2(\mu_i, \nu)
\end{equation*}
with $W_2$ the Wasserstein-2 distance associated with the metric $d$.

From this perspective, the classical $k$-means problem corresponds to the special case $M = 1$, where one seeks to approximate a single empirical distribution $\mu_1$ by a probability measure $\nu$ supported on at most $k$ atoms:
\begin{equation*}
\argmin_{\nu \in \mathcal{P}_k(\Omega)} W_2(\mu_1, \nu),
\end{equation*}
with $\mathcal{P}_k(\Omega)$ the set of probability measures supported on at most $k$ atoms in $\Omega$. In this sense, the solution of a Fr\'{e}chet-$k$ means problem could be seen as a $W_2$ projection of a general probability measure $\mu\in \mathcal{P}(\Omega)$ on $\mathcal{P}_k(\Omega)$. 

\subsection{Optimal Transport for Graph Partitioning} \label{ot_graph}

Optimal Transport (OT) has become a widely used framework in machine learning, with applications ranging from graph matching \parencite{xu2019gromov} and graph dictionary learning \parencite{vincent2021online} to representation learning on structured data. In the context of clustering, it has mainly been employed for clustering sets of graphs \parencite{xu2020gromov}, while node-level clustering within a single graph (referred to hereafter as graph partitioning) has received comparatively less attention.

In this section, we introduce the main notions of OT and its extension to metric spaces through the GW discrepancy \parencite{memoli2011gromov, peyre2016gromov}. We then review existing GW-based approaches for graph partitioning, before presenting our proposed contributions in this setting. 

\subsubsection{Optimal Transport for Graphs}

While classical OT aims to compare probability distributions with supports in the same metric space, \textcite{peyre2016gromov} propose a method for comparing two distance matrices that do not share the same ground space, using the Gromov--Wasserstein (GW) discrepancy \parencite{memoli2011gromov}. This approach seeks to find a matching between these distributions that is as close as possible to an isometry, thus allowing to define a notion of distance between two measured metric spaces. 

This type of approach, along with its variants, may be applied to graphs by representing a graph $G$ as a pair $(\mathbf{R}, \boldsymbol{\mu})$, where $\mathbf{R} \in \mathbb{R}^{N \times N}$ encodes the relationship or connectivity between nodes (adjacency matrix or shortest path distance for example) and $\boldsymbol{\mu} \in \Sigma_N$ (where $\Sigma_N$ is the probability simplex). $\boldsymbol{\mu}$ thus represents the weights or relative importance of the vertices.

\paragraph{Gromov--Wasserstein Discrepancy} Considering two graphs, $G_1 = (\mathbf{R}^{(1)}, \boldsymbol{\mu}^{(1)})$ and $G_2 = (\mathbf{R}^{(2)}, \boldsymbol{\mu}^{(2)})$, with respectively $N$ and $k$ nodes, the Gromow-Wasserstein discrepancy is defined as:

\vspace{-10pt}
\begin{equation*}
    GW_q(\textbf{R}^{(1)}, \boldsymbol{\mu}^{(1)}, \textbf{R}^{(2)}, \boldsymbol{\mu}^{(2)}) = \min_{\substack{\mathbf{T}\mathbf{1}_k = \boldsymbol{\mu}^{(1)} \\ \mathbf{T}^T \mathbf{1}_N = \boldsymbol{\mu}^{(2)}}} \sum_{i,j,l,m} |R^{(1)}_{ij} - R^{(2)}_{lm}|^q T_{il}T_{jm}
\end{equation*}
with $\mathbf{1}_N$ is a column vector of size $N$ where all entries are equal to $1$ and $\mathbf{T} \in \mathbb{R}^{N \times k}_+$ the optimal transport plan, representing the probabilistic matching of nodes.

Note that if $\mathbf{R}$ is a proper distance matrix, then $GW_q$ defines a metric. In contrast, if $\mathbf{R}$ does not satisfy the properties of a distance matrix (for example, if it is an adjacency matrix, which is commonly used for graphs), the resulting quantity is a discrepancy rather than a metric.

\paragraph{Fused Gromov--Wasserstein Discrepancy}

The Gromov--Wasserstein metric focuses solely on structure and is therefore not directly suitable for attributed graphs. \textcite{titouan2019optimal} proposed a new distance for structured data such as attributed graphs, called Fused Gromov--Wasserstein (FGW), which incorporates both topological and feature information:

\begin{align*}
FGW_{q,\alpha}(&\mathbf{R}^{(1)}, \boldsymbol{\mu}^{(1)}, \mathbf{R}^{(2)}, \boldsymbol{\mu}^{(2)}, M_{AB}) \nonumber\\
&= \min_{\substack{\mathbf{T}\mathbf{1}_k = \boldsymbol{\mu}^{(1)} \\ \mathbf{T}^T \mathbf{1}_N = \boldsymbol{\mu}^{(2)}}} 
\sum_{i,j,l,m} \Bigl(
    (1 - \alpha)  d_A(v_i^{(1)}, v_l^{(2)})^q 
    + \alpha |R^{(1)}_{ij} - R^{(2)}_{lm}|^q
\Bigr) T_{il}T_{jm}
\end{align*}

where $M_{AB} = ( d_A(v_i^{(1)}, v_l^{(2)}))_{i,l}$ denotes the $N \times k$ distance matrix between $A$ and $B$, corresponding to the set of attributes of the first and second graphs, respectively. Therefore, the FGW distance seeks the optimal coupling that minimizes a linear combination of the cost of transporting node features from one graph to another and the cost of aligning pairs of nodes according to their respective graph structures.

\paragraph{Semi-relaxed (Fused) Gromov--Wasserstein Discrepancy} \textcite{vincent2021semi} considers that the core optimal transport (OT) assumption of mass conservation may be detrimental for some unsupervised tasks like graph dictionary learning or graph partitioning. To address this, the authors propose relaxing the second marginal and introduce a semi-relaxed Gromov--Wasserstein (srGW) discrepancy:

\begin{equation*}
    \begin{aligned}
        srGW_q(\textbf{R}^{(1)}, \boldsymbol{\mu}^{(1)}, \textbf{R}^{(2)}) & = \min_{\boldsymbol{\mu}^{(2)} \in \Sigma_k} GW_q(\textbf{R}^{(1)}, \boldsymbol{\mu}^{(1)}, \textbf{R}^{(2)}, \boldsymbol{\mu}^{(2)})\\
        & = \min_{\substack{\mathbf{T}\mathbf{1}_k = \boldsymbol{\mu}^{(1)}}} \sum_{i,j,l,m} |R^{(1)}_{ij} - R^{(2)}_{lm}|^q T_{il}T_{jm} 
    \end{aligned}
\end{equation*}

In a similar way, the Fused Gromov--Wasserstein distance was extended in a semi-relaxed setting:

\begin{align*}
srFGW_{q,\alpha}&(\mathbf{R}^{(1)}, \boldsymbol{\mu}^{(1)}, \mathbf{R}^{(2)}, M_{AB}) \\
&= \min_{\substack{\mathbf{T}\mathbf{1}_k = \boldsymbol{\mu}^{(1)}}} 
\sum_{i,j,l,m} \Bigl(
    (1 - \alpha)  d_A(v_i^{(1)}, v_l^{(2)})^q 
    + \alpha \, |R^{(1)}_{ij} - R^{(2)}_{lm}|^q
\Bigr) T_{il}T_{jm}
\end{align*}
Because $\boldsymbol{\mu}^{(2)}$ is optimized within the simplex $\Sigma_k$, the resulting optimal marginal may exhibit sparsity. Consequently, the transport of mass can be concentrated on specific portions of the target structure.

The optimization problem is a non-convex quadratic problem, similar to the one in GW, but with independent linear constraints. \textcite{vincent2021semi} propose solving this problem using a conditional gradient algorithm.

\paragraph{Sensitivity to Initialization}

Both \textcite{chowdhury2021generalized} and \textcite{vincent2021semi} discuss the sensitivity of GW or srGW solvers to initialization, particularly in cases where the target structure provides limited information, such as an identity matrix. In such scenarios, the solver may become trapped in local optima, underscoring the importance of carefully selecting the initialization for the transportation plan. One approach is to use the partitioning results obtained from other existing algorithms (like $k$-means) for initialization.

\subsubsection{Existing Graph Partitioning Methods} \label{existing_gw_methods}

The GW discrepancy and its extension have broad applicability and could be utilized for various tasks, including graph matching and graph coarsening. An additional noteworthy application is graph partitioning, which has been relatively underexplored. The problem no longer consists in comparing two graphs, but rather in matching a source graph $G_s$, represented by $(\mathbf{R}^{(s)}, \boldsymbol{\mu}^{(s)})$, with a much smaller graph $G_t$ of $k$ nodes representing the clusters. The target graph is represented by $(\mathbf{R}^{(t)}, \boldsymbol{\mu}^{(t)})$. $\boldsymbol{\mu}^{(t)}$  require prior knowledge of the relative importance of the classes, information that is often unknown and unconstrained in clustering scenarios. The resulting optimal transport plan provides a soft assignment of each node to the super-nodes of the target graph, thereby defining the final partitions.

\textcite{xu2019scalable} are the first to propose the use of GW-based methods for graph partitioning, thereby raising a number of methodological challenges. Indeed, unlike in graph matching, where both source and target structures are given, in the graph partitioning case the target structure is not predefined. Using the adjacency matrix of the source graph as $\mathbf{R}^{(s)}$, they propose to define the target structure as a disconnected graph with $k$ isolated, self-connected nodes. The central idea is to treat each partition, or subgraph, as a super-node, with the ideal partitioning corresponding to an entirely disconnected target graph. They consider the target graph $G_t$, represented by $(\mathbf{R}^{(t)}, \boldsymbol{\mu}^{(t)})$,
where $\mathbf{R}^{(t)} = \mathbf{I}_k$ and $\mathbf{I}_k$ denoting the identity matrix of size $k \times k$. Another key element that is not specified in graph partitioning is the probability distribution associated with the target graph. They interpolate a node distribution, represented by a vector $\boldsymbol{\mu}^{(t)} \in \Sigma_k$, from the input graph. 

\textcite{chowdhury2021generalized} propose a variant of the previous approach by modifying the source graph representation. Instead of using the adjacency matrix, they consider the (normalized) graph Laplacian $L$ and, more precisely, its heat kernel $K^\lambda = \exp(-\lambda L)$, thereby highlighting the connection between optimal transport and classical spectral partitioning methods. Their experiments show improved performance, but also reveal a strong dependence on the heat kernel scale parameter $\lambda$, which must be carefully tuned. They proposed to tune this scale parameter by maximizing modularity, which results in a higher computational cost.

While the two previous works focus on selecting an appropriate source representation, a key challenge lies in defining the target graph and its associated probability distribution. Using semi-relaxed methods avoids the need to explicitly define the target probability distribution $\boldsymbol{\mu}^{(t)}$. \textcite{vincent2021semi} highlight this advantage by conducting experiments using both adjacency and heat kernel representations, and by comparing the results of semi-relaxed formulation with previous methods that interpolate the target probability distribution, reporting improved performance. Moreover, as discussed earlier, this formulation allows the transported mass to concentrate on fewer than $k$ nodes, thereby alleviating the classical challenge of selecting $k$ in unsupervised clustering.

Regarding attributed graphs, the existing literature remains limited. To the best of our knowledge, no method has so far leveraged the GW discrepancy for attributed graphs partitioning. While graph partitioning with Fused GW has been briefly mentioned \parencite{vayer2020contribution}, the method was originally developed for attributed graphs in a more general setting, and its main applications lie in clustering set of graphs and graph matching.

\subsection{Proposed Contributions for Attributed Graph Partitioning}

As mentioned before, existing methods mainly focus on non-attributed graphs. For graphs with attributes, srGW and srFGW can be interpreted respectively as early-fusion and simultaneous-fusion strategies, in line with the categorization presented in the introduction. Moreover, even in the non-attributed setting, the choice of the target structure (both in terms of graph topology and representation) has not been investigated. In the following section, adaptations to node-attributed graphs are introduced, including novel target constructions, which constitute the core methodological contributions of this work.

\subsubsection{Graph Partitioning with semi-relaxed Gromov--Wasserstein Discrepancy}

\paragraph{\texorpdfstring{Choice of $\mathbf{R}^{(s)}$}{Choice of R}}
Instead of using the adjacency matrix or the heat kernel of the Laplacian, we propose to represent relationships between nodes through a distance matrix. Specifically, we consider $\mathbf{R}^{(s)} = \mathbf{D}_{\alpha}$, the distance matrix that combines topological and attribute-based distances. In the case of non-attributed graphs, the structural distance matrix $\mathbf{D}_{S}$ can be used.

\paragraph{Choice of Target Structure} To the best of our knowledge, no alternative structure than isolated nodes has been proposed so far for partitioning graphs. When employing a normalized distance matrix for $\mathbf{R}^{(s)}$, which encodes both structural and attribute-based relationships between nodes, we propose to define the target structure as $\mathbf{R}^{(t)} = \delta^{(t)} (\mathbf{1}_{k \times k} - \mathbf{I}_k)$ instead of the identity matrix, where $\delta^{(t)}$ denotes the chosen length between target nodes. In this formulation, $\mathbf{R}^{(t)}$ is a distance matrix where intra-group distances are zero, while inter-group distances are strictly positive and can typically be set to the average or maximum distance of the source graph, or even defined through a softmax transformation with temperature optimization.

One may ask if choosing a coarsened graph (a graph with $k$ nodes and the same structural pattern as the source graph) could improve the quality of the partition and reduce the Gromov--Wasserstein distance. Figure~\ref{structure_figure} illustrates examples of possible target graph structures for a chain setting with three groups. The intuition behind this proposition is that a target graph composed of isolated or equidistant nodes may fail to capture the relationships between nodes from different groups in some settings. For example, in a chained graph, the distance between nodes in the first group and those in the last group is large. A $k$ node chain reflects this structure but this inter-group connectivity is not reflected when using a graph of isolated nodes. 

\begin{figure}
    \centering
    
    \subfloat[Source graph.]{
      \begin{minipage}[t]{0.3\textwidth}
        \centering
        \scalebox{0.3}{\input{figures/example_shapes}}
      \end{minipage}
    }
    \subfloat[Target: isolated nodes.]{
      \begin{minipage}[t]{0.3\textwidth}
        \centering
        \scalebox{0.3}{\input{figures/example_shapes2}}
      \end{minipage}
    }
    \subfloat[Target: coarsened graph. \label{target_shapes}]{
      \begin{minipage}[t]{0.3\textwidth}
        \centering
        \scalebox{0.3}{\input{figures/example_shapes3}}
      \end{minipage}
    }
    
    \caption{Illustration of possible target graphs for partitioning a chained source graph with optimal transport.}
    \label{structure_figure}
\end{figure}

\paragraph{Computational Complexity} The proposed version is detailed in Algorithm~\ref{algo_srgw}. All pairwise distances are assumed to be precomputed and are therefore excluded from the computational complexity. The cluster extraction step requires $\mathcal{O}(N k)$ operations. The main computational cost comes from the conditional gradient solver, which requires $\mathcal{O}(N^2k + k^2N)$ operations per iteration \parencite{vincent2021semi}. Let $I$ denoting the maximum number of iterations of the solver, the total computational complexity of the algorithm is $\mathcal{O}(I \cdot (N^2k + k^2N))$, the contributions of the cluster extraction being negligible in comparison.

\begin{algorithm}
    \small
    \caption{semi-relaxed Gromov--Wasserstein for node-attributed graph clustering}
    \begin{algorithmic}
    \STATE \textbf{Input:} An attributed weighted graph $G = (V, E, A)$ characterized by its pair-wise distance matrix $\mathbf{D}_{\alpha} = [d_{\alpha}(v_i, v_j)]_{1 \leq i,j \leq N}$ and its probability distribution $\boldsymbol{\mu}$, $\delta^{(t)}$ the distance between target nodes, $k$ the maximum number of desired subsets and an initial transportation plan
    \STATE Create the target matrix $\mathbf{R}^{(t)} = \delta^{(t)}(\mathbf{1}_{k \times k} - \mathbf{I}_k)$ or consider the representative distance matrix of a coarsened graph
    \STATE Get optimal transport plan $\mathbf{T}^*$ from $srGW(\textbf{D},\boldsymbol{\mu},\mathbf{R}^{(t)})$ with conditional gradient algorithm
    \STATE Get subsets from the last transportation coupling:\\ $C_l = \left\{ i \;\middle|\;
        \forall m \in \{1,\dots,k\},\;
        \begin{cases}
       \mathbf{T}^*_{il} > \mathbf{T}^*_{im} \\
        \text{or } \left( \mathbf{T}^*_{il} = \mathbf{T}^*_{im} \text{ and } l \le m \right)
        \end{cases}
        \right\}$
    \STATE \textbf{return} Subsets $\{C_1, \dots, C_k\}$ where some $C_l$ may be empty.
    \end{algorithmic}
    \label{algo_srgw}
\end{algorithm}

\subsubsection{Graph Partitioning with semi-relaxed Fused Gromov--Wasserstein}

Considering $\mathbf{R}^{(s)} = \mathbf{D}_{\alpha}$ as the representative graph matrix corresponds to an early-fusion strategy, where structural and attribute information are combined prior to applying a partitioning method. FGW (\textcite{titouan2019optimal}), which extends optimal transport to attributed graphs, enables a fused strategy in which partitioning jointly optimizes both structural and attribute components. However, like Gromov--Wasserstein, FGW was not specifically designed for graph partitioning, but rather for more general purposes, and therefore requires certain adaptations for this specific task. 

\paragraph{\texorpdfstring{Choice of $\mathbf{R}^{(s)}$}{Choice of R}} Since attribute distances are specified explicitly in Fused Gromov--Wasserstein, the matrix $\mathbf{R}^{(s)}$ could be chosen as $\mathbf{D_S} = [d_S(v_i, v_j)]_{1 \leq i,j \leq n}$, representing the structural information of the graph, in this case the shortest-path distance matrix.

\paragraph{Optimization Problem}

$M_{AB}$ denotes the $N \times k$ distance matrix between $A$ the set of attributes of the source graph and $B$ the set of attributes of the target graph. In the case of partitioning, however, the attributes of the target graph are, by definition, unknown. In this setting, the optimization problem could be reformulated as:

\begin{align}\label{srfgw_optimization}
srFGW_{q,\alpha}&(\mathbf{R}^{(s)}, \boldsymbol{\mu}^{(s)}, \mathbf{R}^{(t)}, A, d_A) = \min_{\substack{\mathbf{T}\mathbf{1}_k = \boldsymbol{\mu}^{(s)} \\ B \in \mathcal{X}^{k}}} 
\sum_{i,j,l,m} \Bigl(
    (1 - \alpha)  d_A(v_i^{(s)}, b_l)^q 
    + \alpha \, |R^{(s)}_{ij} - R^{(t)}_{lm}|^q
\Bigr) T_{il}T_{jm}
\end{align}

with $B = (b_1, \dots, b_k)$ designs the barycentric attributes of each classes.

It can be observed that, with this formulation, when the problem depends only on the nodes attributes ($\alpha = 0$) and with $q=2$, it reduces to the Fr\'{e}chet $k$-means formulation on attributes:


\begin{align*}
     srFGW_{2,0}(\textbf{R}^{(s)}, \boldsymbol{\mu}^{(s)}, \textbf{R}^{(t)}, A, d_A) & = \min_{\substack{\mathbf{T}\mathbf{1}_k = \boldsymbol{\mu}^{(s)} \\ B \in \mathcal{X}^{k}}} \sum_{i=1}^N \sum_{l=1}^k d_A^2(v_i^{(s)}, b_l) T_{il} \\
     &= \min_{\substack{\boldsymbol{\mu}^{(t)} \in \Sigma_k
     \\B\in\mathcal{X}^{k}}} \min_{\substack{\mathbf{T}\mathbf{1}_k = \boldsymbol{\mu}^{(s)} \\ \mathbf{T}^T \mathbf{1}_N = \boldsymbol{\mu}^{(t)} 
     }}  \sum_{i=1}^N \sum_{l=1}^k d_A^2(v_i^{(s)}, b_l) T_{il}\\
     & = \min_{\nu \in \mathcal{P}_k(\mathcal{X})} W_2^2(\mu^{(s)}, \nu)
\end{align*}
where $\mu^{(1)}$ is the empirical distribution of node attributes, and $\mathcal{P}_k(\mathcal{X})$ the set of probability measures supported on at most $k$ atoms in $\mathcal{X}$.
On the other hand, when focusing solely on the structural part ($\alpha = 1$), the formulation reduces to the semi-relaxed Gromov--Wasserstein method:

\begin{align*}
     srFGW_{q,1}(\textbf{R}^{(s)}, \sum_{i,j,l,m} \boldsymbol{\mu}^{(s)}, \textbf{R}^{(t)}) & = \min_{\substack{\mathbf{T}\mathbf{1}_k = \boldsymbol{\mu}^{(s)}}} |R^{(s)}_{ij} - R^{(t)}_{lm}|^q T_{il}T_{jm} \\
     & = srGW_q(\textbf{R}^{(s)}, \boldsymbol{\mu}^{(s)}, \textbf{R}^{(t)})
\end{align*}

Thus, in the Fused Gromov--Wasserstein framework, $\alpha$ provides a bridge between existing approaches: $\alpha=0$ corresponds to attribute-based Fr\'{e}chet-$k$ means, $\alpha=1$ to structure-based semi-relaxed Gromov--Wasserstein, and values in between yield a combined clustering criterion.

\paragraph{Target Barycenters} \label{propositions}
The optimization problem described in Equation~\ref{srfgw_optimization} also optimizes the attribute barycenters of the target graph. In particular this provides a representative attribute for each cluster.  We propose to initialize barycenters $B$ by computing the weighted barycentric attributes of the classes defined by the initial transportation plan described earlier. The barycenter $b_l$ for each class $C_l$ may either be selected among the nodes of the graph, specifically $b_l \in \argmin_b \sum_{i} d_A(v_i, b)^q T_{il}$, or computed in the full attribute space. The latter option may be more computationally expensive and depends on the ability to compute barycenters with respect to the chosen distance.

Then, the following heuristic could be applied:
\begin{enumerate}
    \item Solve the FGW problem for fixed barycenters $B$.
    \item Update $B$ based on the soft clustering defined by the resulting transportation coupling.
\end{enumerate}

The partitioning algorithm is detailed in Algorithm \ref{algo_srfgw}.

\begin{algorithm}
    \small
    \caption{semi-relaxed Fused Gromov--Wasserstein for node-attributed graph clustering}
    \begin{algorithmic}
    \STATE \textbf{Input:} An attributed weighted graph $G = (V, E, A)$ characterized by its structural distance matrix $\mathbf{D_S}^{(s)}$ and its probability distribution $\boldsymbol{\mu}$, $\delta^{(t)}$ the distance between target nodes, the maximum number of desired subsets $k$, an initial transportation plan $\mathbf{T}^0$ and a maximal number of iterations
    \STATE \textbf{Initialization:}
        \STATE Create the target matrix $\mathbf{D}_S^{(t)} = \delta^{(t)}(\mathbf{1}_{k \times k} - \mathbf{I}_k)$ or consider the structural distance matrix of a coarsened graph
        \STATE n = 0
        \STATE $\mathbf{T}^n = \mathbf{T}^0$
    \REPEAT
        \STATE Compute $B^n = (b_1^n, \dots, b_k^n)$ the weighted barycentric attributes of each subset
        \STATE  Compute $\mathbf{M} = \left[ d_A(v_i, b_j) \right]_{\substack{1 \leq i \leq N \\ 1 \leq j \leq k}}$, the distance matrix between the node attributes and the barycentric attributes of each class.
        \STATE Get optimal transport plan $\mathbf{T}^{n+1}$ from $srFGW(\mathbf{D}_S,\boldsymbol{\mu},\mathbf{D}_S^{(t)},\mathbf{M})$
        \STATE $k = \#\{ l \in \{1, \dots, k\} \mid C_l \neq \emptyset \}$
        \STATE $n = n+1$
        \UNTIL{$T^n = T^{n-1}$ \textbf{ or } $n >$ \textit{maximal number of iterations}}
        \STATE (optional) Get subsets from the last transportation coupling:\\ $C_l^n = \left\{ i \;\middle|\;
        \forall m \in \{1,\dots,k\},\;
        \begin{cases}
       \mathbf{T}^n_{il} > \mathbf{T}^n_{im} \\
        \text{or } \left( \mathbf{T}^n_{il} = \mathbf{T}^n_{im} \text{ and } l \le m \right)
        \end{cases}
        \right\}$
    \STATE (optional) Update the barycentric attributes according to the last hard clustering step
    \STATE \textbf{return} Subsets $\{C_1, \dots, C_k\}$ where some $C_l$ may be empty, and their corresponding barycentric attributes.
    \end{algorithmic}
    \label{algo_srfgw}
\end{algorithm}
\begin{proposition}\label{prop:loss_decreasing} Let $\mathcal{L}(T, B)$ denote the semi-relaxed Fused Gromov--Wasserstein loss 
for a transportation plan $T$ and barycentric attributes $B$. 
\begin{equation*}
    \mathcal{L}(T, B) =
\sum_{i,j,l,m} \Bigl(
    (1 - \alpha)  d_A(v_i^{(s)}, b_l)^q 
    + \alpha \, |R^{(s)}_{ij} - R^{(t)}_{lm}|^q
\Bigr) T_{il}T_{jm}
\end{equation*}
If $(T^n,B^n)$ are the iterates produced by Algorithm~\ref{algo_srfgw}, 
then the sequence $(\mathcal{L}(T^n,B^n))_{n\geq 0}$ is monotonically non-increasing. 
More precisely, at each iteration $n$, it holds that
\begin{equation*}
    \mathcal{L}(T^{n+1}, B^{n+1}) \leq \mathcal{L}(T^{n+1}, B^n) \leq \mathcal{L}(T^n, B^n)
\end{equation*}
\end{proposition}

The proof is deferred to Appendix~\ref{app:proof_loss_decreasing}.

\begin{proposition}\label{prop:hard_clustering}

Let $(T, B)$ be a solution obtained by Algorithm \ref{algo_srfgw}. Let $\tilde T$ be another transport plan obtained from $T$ by any projection or modification. For each cluster $C_l:=\{i:\ \tilde T_{il}>0\}$, let the associated barycenter be recomputed as
\begin{equation*}
    \tilde b_l \in \argmin_b \sum_{i\in C_l} \mu_i d_A(v_i,b)^q,
\end{equation*}

Define
\begin{equation*}
D_A:=\max_{\substack{1\le i\le N \\ 1\le l\le k}} d_A(v_i,b_l),
\qquad
D_S:=\max_{\substack{1\le i,j\le N \\ 1\le m,l\le k}}|R^{(1)}_{ij}-R^{(2)}_{lm}|,
\end{equation*}

with respect to the original barycenters $B$.

Then the loss increase induced by replacing $(T,B)$ with $(\tilde T,\tilde B)$ satisfies

\begin{equation*}
\mathcal{L}(\tilde T, \tilde B)-\mathcal{L}(T,B)
 \le
((1-\alpha)D_A^q+2\alpha D_S^{\,q})\ \sum_i \sum_l |T_{il}-\tilde T_{il}|.
\end{equation*}
\end{proposition}
The eventual increase in loss is proportional to the total deviation between the two couplings, with proportionality constants depending only on the maximal attribute and structural dissimilarities of the data.

The proof is deferred to Appendix~\ref{app:hard_clustering}.

\begin{remark}
A notable special case is the soft-to-hard projection. Let $T^{\mathrm{soft}}$ denote a transportation plan yielding a soft clustering, and let $B^{\mathrm{soft}}$ denote its associated barycenters. For each row $i\in\{1,\dots,n\}$, a hard projection is done based on the following deterministic tie-break rule: 
\begin{equation*}
    T^{\mathrm{hard}}_{il}=
    \begin{cases}
    \mu_i,& \text{if} \quad l=\min \{m\in\{1,\dots,k\}:\ T_{il}=\max_{1\le r\le k}T^{\mathrm{soft}}_{ir}\},\\
    0,& \text{otherwise.}
    \end{cases}
\end{equation*}

Let $B^{\mathrm{hard}}=( b_1^{\mathrm{hard}},\dots, b_k^{\mathrm{hard}})$ be any choice of barycenters recomputed on the hard clusters $C_l$.
Then the total deviation between the soft and hard assignment matrices becomes:
$$ \sum_i \sum_l |T^{\mathrm{soft}}_{il}-T^{\mathrm{hard}}_{il}|=2\sum_{i=1}^n
(\mu_i-m_i^{soft}),$$
with $m_i^{\mathrm{soft}}=\max_{l}T_{il}^{\mathrm{soft}}$. This difference is maximal when the soft clustering assigns each node to all classes with uniform weight. 
Thus, the maximal loss augmentation that could be induced by the optional hard clustering step at the end of Algorithm \ref{algo_srfgw} satisfies:
\begin{equation*}
\mathcal{L}(T^{\mathrm{hard}},B^{\mathrm{hard}})-\mathcal{L}(T^{\mathrm{soft}},B^{\mathrm{soft}})
 \le
2((1-\alpha)D_A^q+2\alpha D_S^{\,q})\ (1-\frac{1}{k}).
\end{equation*}

\end{remark}
%
%

\paragraph{Computational Complexity}
Following the previous setting, pairwise distances are assumed to be precomputed and are excluded from the complexity evaluation. When barycenters are selected among the nodes, their computation requires $\mathcal{O}(N^2)$ operations per iteration of the main algorithm. Each iteration of the conditional gradient solver for srFGW requires $\mathcal{O}(N^2k + k^2N)$ operations and is repeated maximum $I$ times per main iteration. Finally, the cluster extraction step is performed once at the end, with a cost of $\mathcal{O}(N^2k + k^2N)$. Letting $n$ denote the number of iterations of the main algorithm, the total computational complexity is $\mathcal{O}(n(N^2 + I \cdot (N^2k + k^2N)))$ with the contribution of the final cluster extraction being negligible compared to the barycenter and solver computations. When accounting for both the main iterations and the solver iterations, this method emerges as the most computationally expensive among the approaches considered.

\subsubsection{Feature Map Representation}
Feature mapping is a technique used in machine learning to transform input data into a format that can be analyzed more easily. It is often applied when dealing with complex structured data like texts, images, sounds, etc. For example, texts are sometimes  represented as frequency vectors over words. Mapping the data into a larger dimensional space may also help uncover underlying patterns that are harder to detect into the original space. This is the main idea behind kernel methods, although kernel methods do not necessarily need an explicit form for the feature map. 

Generally, a feature map is a function $\phi:\mathcal{X}\to \mathcal{H}$, going from an input set $\mathcal{X}$ to a representative space $\mathcal{H}$. While $\mathcal{X}$ could practically be any non-empty set, with no additional mathematical structure, the representative space $\mathcal{H}$ is often chosen as a vector (or a Hilbert space for kernel based approaches), where the classical linear statistical methods can be applied. 


Feature maps may also be used for graphs. The vertices of a graph $G=(V,E)$ could be represented in an Euclidean space using the structure of $G$ in multiple ways. For example, one could define $\phi:V\to\mathbb{R}^N$, as $\phi(v_i)=(a_{i1},\ldots,a_{in})$, where the $a_{ij}$'s encode the adjacency information ($a_{ij}=1$ if there is an edge between $v_i$ and $v_j$ and $0$ otherwise). The feature map $\phi$ may also be defined using $d$, the geodesic distance of the graph, by setting $\phi(v_i)=(\ldots,d(v_i,v_j),\ldots)$.

\cite{li2017locally} apply this idea of feature representation in order to construct an adaptive weighted $k$-means algorithm for attributed graphs. They suggest representing each node of a graph using a linear combination of two maps: a map $\phi_S:V\to\mathbb{R}^p$ encoding structural information and a map $\phi_A:A\to\mathbb{R}^p$, encoding information related to attributes ($p$ not necessarily equal to the number of nodes in the graph). Thus, each attributed node $(v_i,a_i)$ is mapped to a vector $\varphi_{\alpha}(v_i,a_i)=\alpha_i\phi_S(v_i)+(1-\alpha_i)\phi_A(a_i)\in \mathbb{R}^p$. Then they propose to cluster the nodes using a $k$-means procedure on $\{ \varphi_{\alpha}(v_i,a_i)\ | \ i=1\ldots N\}$, optimizing the $k$-means criteria both over the barycenters' positions (as usually), and over the weights $\alpha=(\alpha_1,\ldots,\alpha_N)\in [0,1]^N$, that model the importance given to the structural information for each node. 
They illustrate the interest of this approach by evaluating the algorithm's performance on real data sets (graphs with textual attributes), using a structural representation based on the adjacency matrix.

The idea of feature map representation may also be applied for the OT-based methods described above. While optimizing the weights $\alpha$, or even exploring other choices of representations could be an interesting lead, it is out of scope for this paper and it is left for future research. Here, a feature map representation is applied as a preprocessing step. In the following, the use of an embedding for each method is described. 
\paragraph{srGW Clustering with Embedding} In the case of a non-attributed graph, we propose to map each node to the vector $\phi_S(v_i)=(\ldots,d_S(v_i,v_j),\ldots)\in\mathbb{R}^N_+$ and then construct the associated distance matrix $(\mathbf{D}_{S}^{(1)})_{i,j}=\|\phi_S(v_i)-\phi_S(v_j)\|_2$. A GW based clustering can be obtained by applying Algorithm \ref{algo_srgw} to the matrix $\mathbf{D}_{S}^{(1)}$. For an attributed graph, each attributed node $(v_i,a_i)$ could be mapped to a vector $\varphi_{\alpha}(v_i,a_i)=\alpha(\ldots ,d_S(v_i,v_j),\ldots)+(1-\alpha)(\ldots,d_A(a_i,a_j),\ldots)\in \mathbb{R}^N_+$. Similarly to the non-attributed case, Algorithm \ref{algo_srgw} can then be applied to the matrix $\mathbf{D}_{\alpha}^{(1)}$, which contains the pairwise Euclidean distances between the embedded points,  $(\mathbf{D}_{\alpha}^{(1)})_{i,j}=\|\varphi_{\alpha}(v_i,a_i)-\varphi_{\alpha}(v_j,a_j)\|_2$. 

\paragraph{srFGW Clustering with Embedding} Since the fused version treats the attributes separately and in a way, it deals with them directly in their ambient space $\mathcal{X}$, we propose to use an embedding only for the structural information. To each vertex $v_i$, we associate a vector $\phi_S(v_i)$ in order to construct the associated distance matrix $(\mathbf{D}_{S}^{(1)})_{i,j}$. Algorithm \ref{algo_srfgw} can now be applied by replacing the initial structural distance matrix $\mathbf{D}_{S}$ by $\mathbf{D}_{S}^{(1)}$.

%% file: figures/example_shapes.tex
\begin{tikzpicture}[
    every node/.style={circle, draw=black, thick, minimum size=8mm, inner sep=0pt},
    cluster1/.style={fill=dkgreen!30},
    cluster2/.style={fill=purple!30},
    cluster3/.style={fill=orange!30},
    arrow/.style={-Latex, thick},
]

\node[] (1) at (0,0) {};
\node[] (2) at (-1,1) {};
\node[] (3) at (0,1) {};
\node[] (4) at (1,1) {};
\node[] (5) at (0,2) {};

\node[] (6) at (0,-2) {};
\node[] (7) at (0,-3) {};
\node[] (8) at (0,-4) {};
\node[] (9) at (1,-3) {};
\node[] (10) at (-1,-3) {};

\node[] (11) at (0,-6) {};
\node[] (12) at (0,-7) {};
\node[] (13) at (0,-8) {};
\node[] (14) at (1,-7) {};
\node[] (15) at (-1,-7) {};

\draw (1) -- (2);
\draw (1) -- (3);
\draw (1) -- (4);
\draw (2) -- (5);
\draw (2) -- (3);
\draw (3) -- (5);
\draw (4) -- (3);
\draw (4) -- (5);

\draw (1) -- (6);

\draw (6) -- (7);
\draw (7) -- (8);
\draw (8) -- (9);
\draw (9) -- (7);
\draw (9) -- (6);
\draw (7) -- (10);
\draw (8) -- (10);
\draw (10) -- (6);

\draw (8) -- (11);

\draw (11) -- (12);
\draw (12) -- (13);
\draw (13) -- (14);
\draw (14) -- (12);
\draw (14) -- (11);
\draw (12) -- (15);
\draw (13) -- (15);
\draw (15) -- (11);

\end{tikzpicture}

%% file: figures/example_shapes2.tex
\begin{tikzpicture}[
    every node/.style={circle, draw=black, thick, minimum size=8mm, inner sep=0pt},
    arrow/.style={-Latex, thick},
]

\node[] (1) at (0,0) {};

\node[] (2) at (0,-4) {};

\node[] (3) at (0,-8) {};

\draw[->] (1) edge[loop above] ();
\draw[->] (2) edge[loop above] ();
\draw[->] (3) edge[loop above] ();

\end{tikzpicture}

%% file: figures/example_shapes3.tex
\begin{tikzpicture}[
    every node/.style={circle, draw=black, thick, minimum size=8mm, inner sep=0pt},
    arrow/.style={-Latex, thick},
]

\node[] (1) at (0,0) {};

\node[] (2) at (0,-4) {};

\node[] (3) at (0,-8) {};

\draw (1) -- (2);
\draw (2) -- (3);

\end{tikzpicture}

%% file: sections/3_simulations.tex
\FloatBarrier
\section{Evaluation on Synthetic Data} \label{sec : Results}

This section presents several experimental studies on synthetic data. First, we investigate the impact of several factors (such as initialization choice, the use of embeddings within the methods, and the influence of the target choice) on non-attributed graphs, since only the graph topology affects these outcomes. In a second step, the methods are compared on attributed graphs to assess their sensitivity to both structural and attribute-related noise.

Note that, in the spirit of reproducible research, code is available on a \href{https://github.com/KetsiaGuichard/ot-based-graph-partitioning/}{public Github repository}. The implementation is Python-based and relies on the POT toolbox \parencite{flamary2021pot}.

\subsection{Experiments on Non-Attributed Graphs}

For the simulation, a non-attributed graph $G = (V,E)$ is considered, where $V$ is the set of $N$ nodes ($|V| = N$), and $E$ denotes the set of undirected edges, with no self-loops. Focus on non-attributed graphs simplify the analysis, avoiding the need to define attribute-based similarities between groups.

This setting also allows us to assess the impact of initialization choices, compare embedded and non-embedded methods, and isolate the effect of the target graph structure on partitioning performance within a controlled environment.

To facilitate the extension of our conclusions to the broader setting of attributed graphs, graph structures are represented through distance matrices when applying the proposed algorithm in the simulation study. Our methods (semi-relaxed Gromov–Wasserstein clustering based on distances with equidistant nodes as target, and its embedding-based variant) are compared with semi-relaxed Gromov–Wasserstein clustering based on adjacency, Fr\'{e}chet $k$-means and spectral clustering based on adjacency. 

\subsubsection{Graph Generation}
Simple undirected graphs without attributes are simulated, with predefined structural patterns. These graphs consist of well-separated communities, characterized by high intra-group edge density and relatively sparse inter-group connectivity. To generate such structures, Stochastic Block Model (SBM) is employed, a classical generative model for random graphs that emphasizes community organization. In the SBM, $N$ nodes are partitioned into $k$ groups, and edges are independently sampled between node pairs according to a group-dependent probability. More precisely, the model defines a probability matrix $\mathbf{P} = [p_{rs}] \in [0, 1]^{k \times k}$, where $p_{rs}$ denotes the probability of an edge between a node in group $r$ and a node in group $s$. This matrix, referred to as the block matrix or community connection matrix, typically features larger diagonal values (i.e., $p_{rr} > p_{rs}$ for $r \neq s$), promoting dense intra-community connections and sparse inter-community links.

In the simple case where all groups are connected to each other (the fully connected setting), the block probability matrix is generated from a $k \times k$ matrix $\mathbf{P}$ whose diagonal entries are equal to $p_{\mathrm{intra}}$ and whose off-diagonal entries are equal to $p_{\mathrm{inter}}$, with both parameters belonging to $[0,1]$. The quantity $\rho = \frac{p_{\mathrm{inter}}}{p_{\mathrm{intra}}}$ therefore represents the relative strength of inter-group connectivity with respect to intra-group connectivity.  Denoting by $\mathbf{I}$ the $k \times k$ identity matrix and by $\mathbf{1}$ the $k \times k$ matrix of ones, the block probability matrix is given by:

\begin{equation*}
    \mathbf{P}=p_{\mathrm{intra}} \cdot \mathbf{I}+p_{\mathrm{inter}} \cdot (\mathbf{1} - \mathbf{I}).
\end{equation*}

In our experiments, and in order to obtain degree distributions comparable to those observed in road networks, where node degrees are typically small (ranging from 1 to 5), $p_{\mathrm{inter}}$ is fixed to relatively small value ($0.08$), while $\rho$ varies from $0.01$ to $0.20$. These parameter choices generate networks with increasingly weak community structure as $\rho$ increases, while remaining within the regime where community recovery is theoretically possible \parencite{abbe2018community}.

\begin{figure}
\centering
\includegraphics[width=\textwidth]{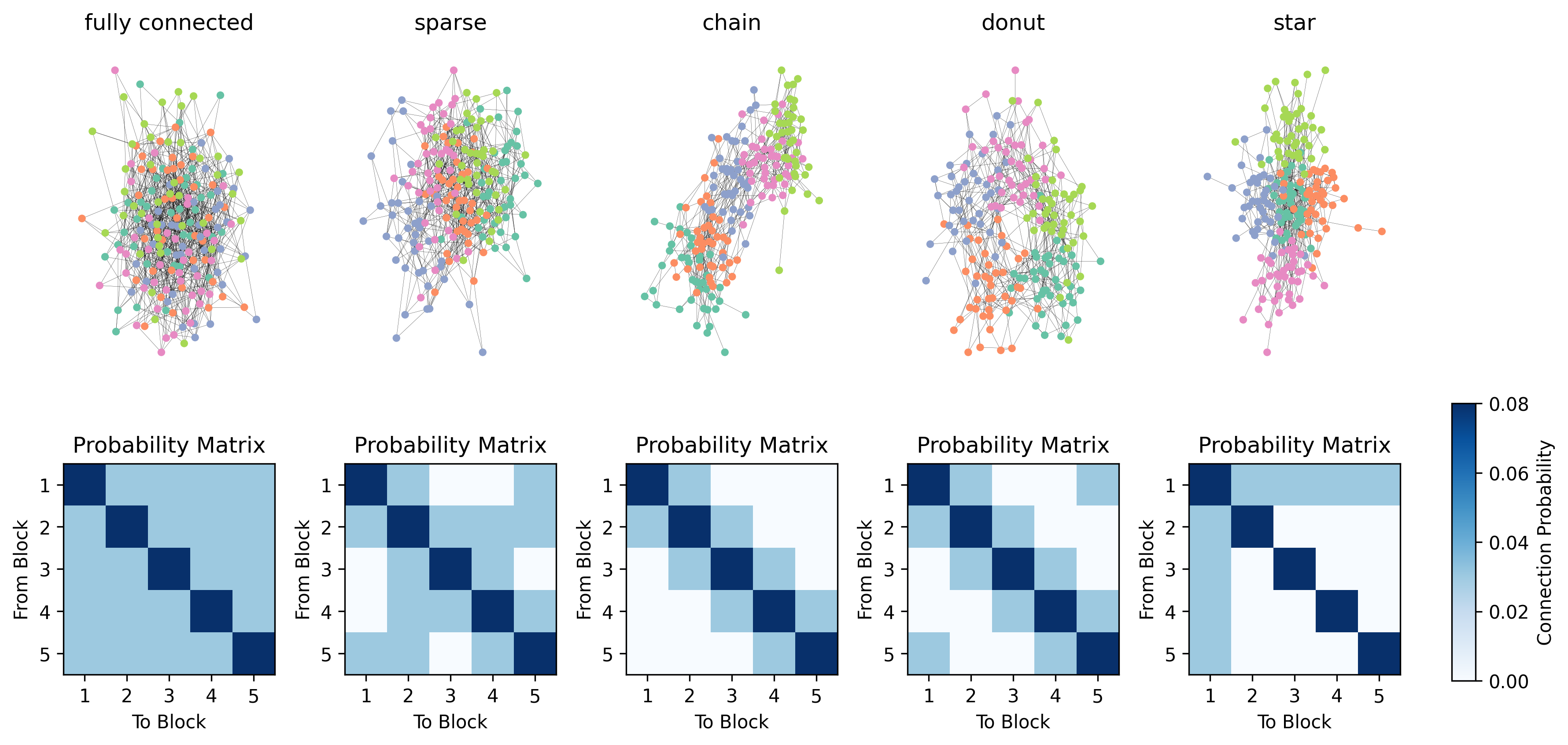}
\caption{Illustration of different graph shapes and their respective community connection matrices in the the stochastic block model (SBM) framework ($p_{\mathrm{intra}}= 0.08, \rho=0.2$).}
\label{heatmaps_graphs_highres}
\end{figure}

More complex graph topologies with well-separated groups, such as sparse, chain, donut, or star-shaped graphs, are also considered. This is particularly relevant when studying the effect of the target graph structure, but also allows us to evaluate the influence of the graph's topology on the  performance of the methods. These structures are illustrated in Figure~\ref{heatmaps_graphs_highres}.
To generate them, additional constraints are applied to the block matrix, such as enforcing sparsity (e.g., a fixed proportion of zero entries) or restricting connections to specific group pairs, according to the desired topology.

\paragraph{Graph Distance} For the structural distance $d_S$, the geodesic distance is considered, represented by the length of the shortest path between two nodes, normalized:     $\tilde{d}_S(v_i, v_j) = \frac{d_S(v_i, v_j)}{\max(d_S)}$.

\subsubsection{Evaluation} 

The methods are evaluated on graphs composed of 200 nodes, distributed across 5 communities of comparable size, using Monte Carlo simulations with 100 repetitions per setting.

The ground truth partition generated by the SBM is compared with the partitions produced by the methods. In cases where the transportation plan assigns a node to multiple classes, the node is assigned to the node receiving the largest amount of transported mass (or the first one in case of ties), resulting in a hard clustering. Different strengths of community structure are tested and the performance is assessed using a widely recognized external clustering evaluation metric: the Adjusted Rand Index (ARI, \textcite{hubert1985comparing}). The ARI adjusts the Rand Index (RI) for chance grouping and quantifies the similarity between two partitions by measuring the proportion of node pairs that are consistently placed either in the same cluster or in different clusters across both partitions.

\subsubsection{Initialization Choice}

The sensitivity of our methods to initialization naturally raises the question of how to select the initial centroids or the initial transport plan. Three initialization strategies are considered: randomly selecting $k$ centroids, applying $k$-means++ on $V$, or applying $k$-means++ to the embedded nodes. The latter approach will hereafter be referred to $k$-means++ on $D$.

Table~\ref{init_impact} shows the impact of the three initialization strategies on a sparse SBM graph with $p_\textrm{intra}=0.08$ and $\rho=0.1$ for Fr\'{e}chet $k$-means, srGW with equidistant nodes, and its embedded variant, considering both average and maximum distances between target nodes. The $k$-means++ on $D$ strategy achieved the best results across almost all methods, with the improvement being particularly notable for GW-based approaches with maximum distances. In the remainder of this paper, all these methods are initialized using the $k$-means++ on $D$ strategy.

\begin{table}
\caption{Performance comparison (average ARI) of initialization strategies on a SBM graph $(p_\textrm{intra}=0.08, \rho=0.1)$ with a sparse structure.}
\label{init_impact}
\centering
\begin{tabular}{@{}lccccc@{}}
\hline
& \makecell{Frechet \\ $k$-means} 
& \makecell{srGW \\ (mean)} 
& \makecell{srGW \\ (max)} 
& \makecell{Embedded \\ srGW (mean)} 
& \makecell{Embedded \\ srGW (max)} \\\hline
Random & 0.378  & \textbf{0.842}& 0.250 & 0.780 & 0.349 \\
$k$-means++ on $V$ & 0.393 & 0.832 & 0.259 & 0.781 & 0.352 \\
$k$-means++ on $D$ & \textbf{0.408} & 0.836 & \textbf{0.273} & \textbf{0.784} & \textbf{0.370} \\
\hline
\end{tabular}
\end{table}

Note that this $k$-means++ strategy can also be applied multiple times, after which the initialization minimizing the optimal transport cost is selected. However, for computational cost reasons, this strategy is not applied in this simulation section.

\subsubsection{Influence of the Source and Target Choices}

\paragraph{Source Structure}
As discussed before, to the best of our knowledge, graphs have so far only been represented for partitioning in the literature through either their adjacency matrix or a heat kernel. Since the latter approach requires the optimization of a scale parameter, it is not considered in the scope of this paper. We compare the results obtained using the classical adjacency matrix representation with those obtained using the distance matrix proposed in the methodological section.

\paragraph{Target Structure} In the case where the graph is represented by its adjacency matrix, the only target graph structure that has been proposed and tested for clustering so far corresponds to a disconnected graph with $k$ isolated nodes. For our distance-matrix-based alternative, three types of target structures are considered: two versions of equidistant nodes with different inter-node distances (the average distance and the maximum distance in the source graph), and a summarized representation of the input graph. This latter choice is motivated by the hypothesis that using a target structure that more closely approximates the actual topology of the graph may yield better performance than the standard baseline.

\begin{figure}
\centering
\includegraphics[width=\textwidth]{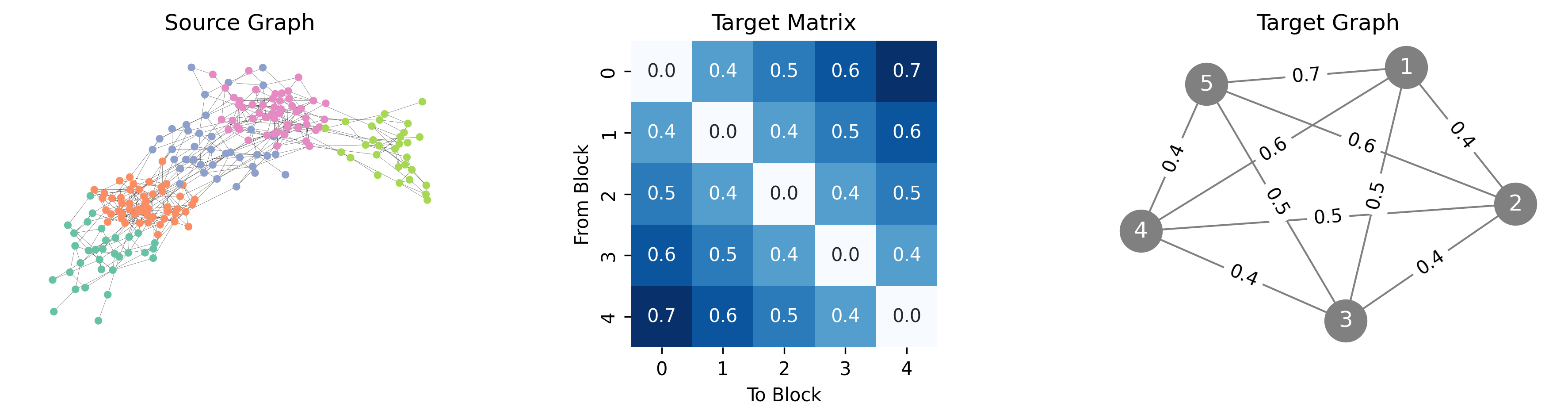}
\caption{Example of a generated chain graph, its median shortest-path distance matrix between groups, and the resulting target graph.}
\label{target_graph}
\end{figure}

For the coarsened graph version, since the graphs are synthetically generated, the ground-truth partition is known, allowing the construction of a meaningful target structure. In particular, the SBM graph is leveraged, and the median shortest-path distance between groups is computed to define the target matrix. An illustration of this principle is provided in Figure~\ref{target_graph}.

\paragraph{Results} The partitions produced by the different semi-relaxed Gromov--Wasserstein variants are compared with the ground-truth partition generated by the SBM. Since the graphs considered in this experiment do not include node attributes, only semi-relaxed Gromov--Wasserstein methods are evaluated, as the semi-relaxed Fused Gromov--Wasserstein (srFGW) framework is not applicable in this setting.

\begin{figure}
    \centering
    \begin{subfigure}{1\textwidth}
        \centering
        \includegraphics[width=\textwidth]{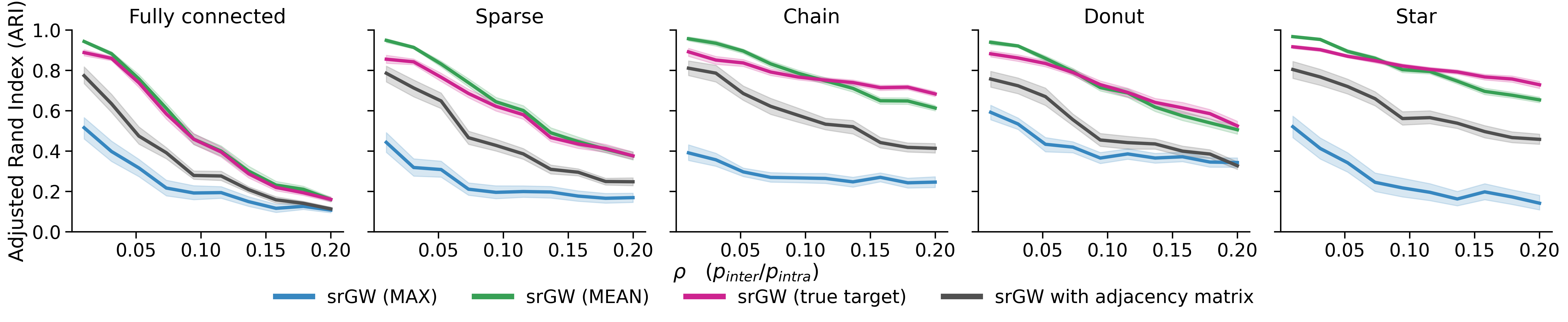}
        \caption{Comparison of semi-relaxed Gromov--Wasserstein methods across different source and target graph representations.}
        \label{srGW_comparison_nonattributed}
    \end{subfigure}
    
    \vspace{0.5cm}
    
    \begin{subfigure}{1\textwidth}
        \centering
        \includegraphics[width=1\textwidth]{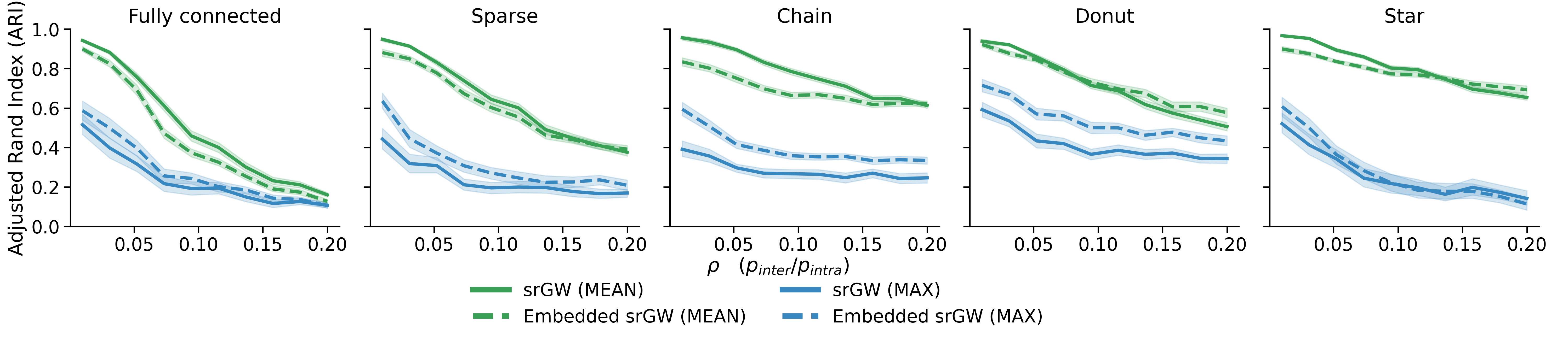}
        \caption{Comparison of embedded and non-embedded methods.}
        \label{embedded_impact_nonattributed}
    \end{subfigure}
    
    \vspace{0.5cm}
    
    \begin{subfigure}{1\textwidth}
        \centering
        \includegraphics[width=1\textwidth]{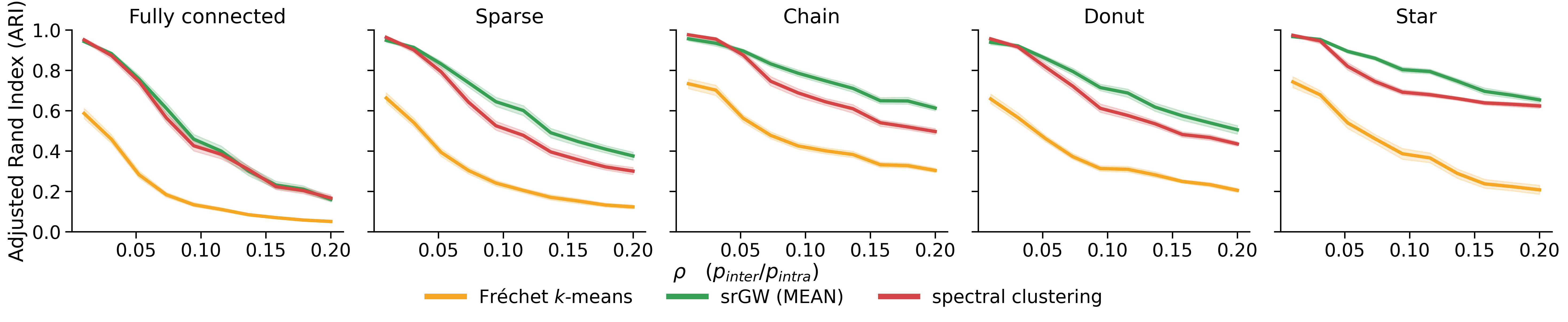}
        \caption{Comparison of the best-performing semi-relaxed Gromov--Wasserstein method with non-Gromov-based methods.}
        \label{comparison_other_methods_nonattributed}
    \end{subfigure}
    \caption{Comparison of partitioning performance (ARI) of methods across varying levels of non-attributed graph perturbation and various shapes.}
\end{figure}

Figure~\ref{srGW_comparison_nonattributed} highlights the benefit of the proposed distance-based formulation. The method of \textcite{xu2019scalable}, which relies on the adjacency matrix as the source representation and the identity matrix as the target, consistently yields lower clustering performance. In contrast, representing graph structure through the distance matrix and using a target matrix of equidistant nodes (with the common distance set to the mean shortest-path distance in the source graph) significantly improves partition recovery.

This choice achieves performance comparable to that of a coarsened graph used as the target. While mean-equidistant nodes perform slightly better when intra- and inter-group probabilities are well separated, the coarsened target becomes advantageous as the distinction between groups decreases. This remark is particularly valid in settings where the graph exhibits strong structural patterns, such as chain, donut or star configurations. However, the coarsened target relies on strong prior knowledge of the graph structure, while providing only limited improvements over the much simpler equidistant-node target.

\begin{figure}
  \centering
  \subfloat[Distance matrices of a chain-shaped SBM and its noisy variant. \label{noisy:matrices}]{
    \begin{minipage}[c]{0.55\textwidth}
      \centering
      \includegraphics[width=0.95\linewidth]{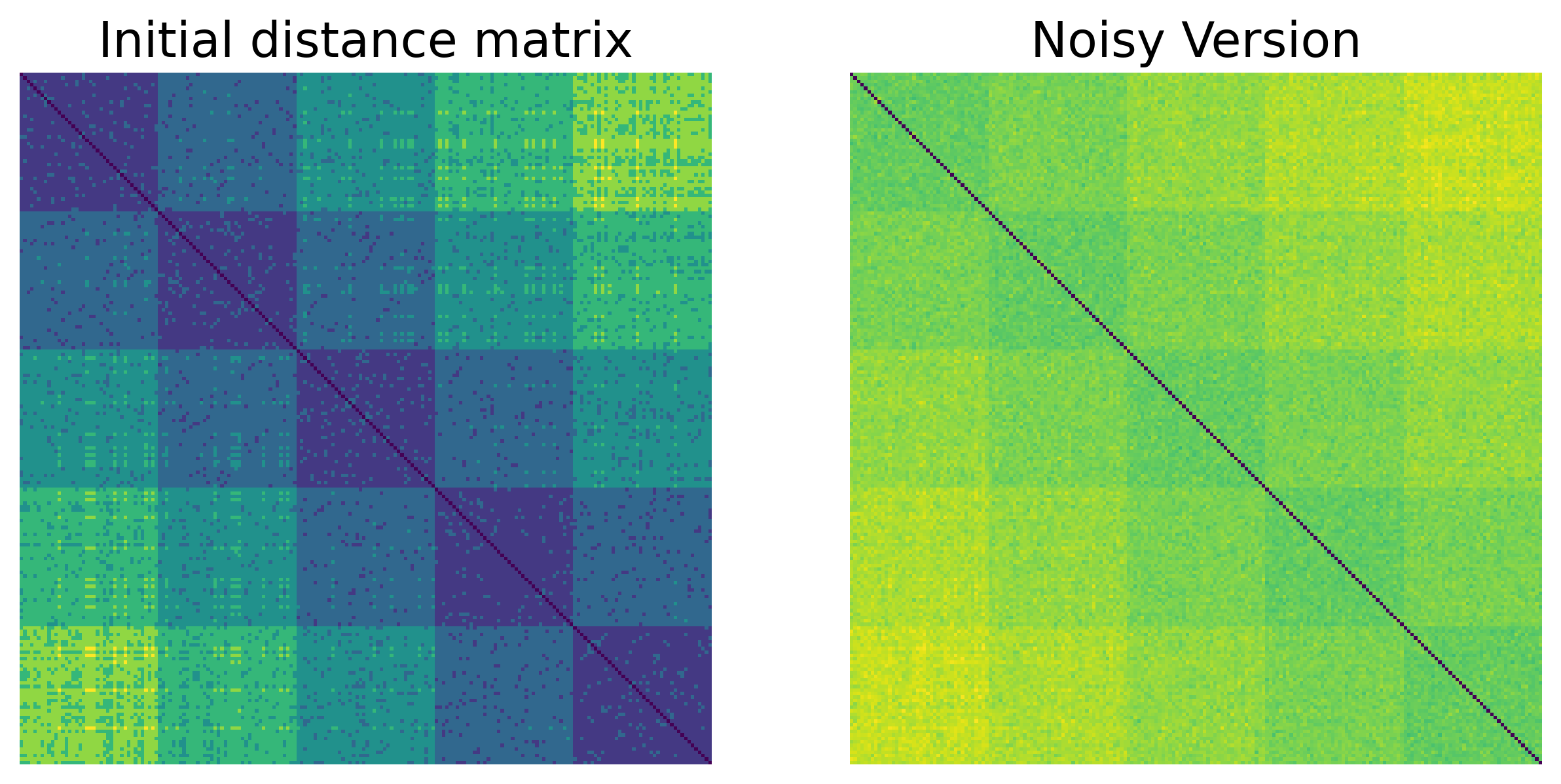}%
    \end{minipage}%
  }
  \hfill
  \subfloat[Mean ARI for each target and distance matrix. \label{noisy:table}]{
    \begin{minipage}[c]{0.40\textwidth}
      \centering
      \begin{tabular}{lcc}
      \hline
       & srGW (max) & srGW (mean) \\
      \hline
      Initial & 0.864 & \textbf{0.985} \\
      Noisy   & \textbf{0.977} & 0.952 \\
      \hline
      \end{tabular}
    \end{minipage}%
  }
  \caption{Target performance in noisy conditions.}
  \label{noisy}
\end{figure}

The choice of the distance between equidistant nodes could for example be made by selecting the version with the smallest GW criterion. Even though this is not the case in our current experiments, it is observed that in some settings one choice is better suited than the other. For highly contrastive distance matrices (graphs with large differences between intra- and inter-group distances) the average strategy performs better than using the maximum value. Indeed, in this case, the average distance better reflects inter-group distances. Conversely, when intra- and inter-group distances are close, using the maximum value as the distance between target nodes may help to distinguish the groups. This idea is illustrated in Figure \ref{noisy}, which shows the performance of two distance choices between target nodes for a chain SBM graph and its noisy version (obtained by adding a Gaussian noise to the initial distance matrix).

\paragraph{Comparison With Embedded Methods}

For non-attributed graphs, Figure~\ref{embedded_impact_nonattributed} shows that embedding-based representations outperform their non-embedded counterparts when the maximum distance is used in the target structure. When the average distance is used, they generally achieve lower performance, especially in easier settings, where groups are well separated (small $\rho$). However, they match or outperform their counterparts in the most challenging settings, characterized by weaker structural signals, particularly when the graph exhibits strong topological patterns such as donut or star structures. Overall, in these latter settings, embedding-based representations appear to be more robust to perturbations of the graph structure. The performance of embedded representations will be further discussed in the attributed graph simulation study presented in Section~\ref{fgw_section}.

\subsubsection{Comparison with Non-Gromov-Wasserstein Methods}

So far, semi-relaxed Gromov--Wasserstein (srGW) methods have only been compared against each other to identify the best source and target representations. Figure~\ref{comparison_other_methods_nonattributed} provides a comparison with non-GW methods, namely Fr\'{e}chet $k$-means, another distance-based method previously introduced in Section~\ref{kmeans_section}, and spectral clustering based on the graph Laplacian, a standard baseline for this type of task.

Semi-relaxed GW with a distance-matrix source and a mean-equidistant target largely outperforms Fr\'{e}chet $k$-means, which yields relatively poor results, despite its appeal as a simple, computationally efficient, and easily interpretable baseline. Compared with spectral clustering, it achieves similar performance in the fully connected setting but better performance in other regimes, particularly when the task becomes more challenging, when intra- and inter-group connection probabilities are closer.

\subsection{Performance Comparison on Synthetic Attributed Graph}

In addition to the synthetic non-attributed graphs, node attributes are also simulated to evaluate both the performance and computational efficiency of the proposed methods. 

In a first step, attributes are generated to reflect those commonly found in transportation applications, such as functional data for traffic flow curves and histograms for speed distributions. However, the proposed framework is not restricted to these attribute types and can be applied more generally, provided that appropriate distance metrics and barycenter operators are available.

To investigate the impact of graph topology more precisely, a second simulation setting is also considered. In this case, attribute dissimilarity matrices are generated directly so that the attribute structure follows a prescribed topology (for instance a chain graph). This construction provides greater control over the relationship between structural and attribute information and makes it possible to assess the methods under well-defined topological patterns.

\subsubsection{Generation of Transportation-Like Attributes}

The focus here is on fully connected graphs, with attributes generated in alignment with the underlying SBM structure: each community shares a common attribute model, and intra-group variability is introduced through controlled perturbations.

\paragraph{Functional Data Generation}
Basis splines (or B-splines) will be used to generate the different class models and their noisy version. B-splines are flexible basis commonly used for curve fitting in data analysis \parencite{ramsay2005functional}. They are piece-wise polynomial functions joint at different knots. Let us consider $m$ fixed knots and $p$ the degree of polynomial functions.
The associated basis functions are denoted $B_{l,p}(x)$, with $1 \leq l \leq m+p-2$. A spline $S(x)$, spanned by this basis, can then be expressed as: 
\begin{equation*}
    S(x) = \sum_{l} \theta_l B_{l,p}(x) \textrm{, with $\theta_l$ the coefficients associated with each basis function.}
\end{equation*}
For each group $j \in \{1, \dots, k\}$, the coefficients $\boldsymbol{\theta}^{(j)} = (\theta_1^{(j)}, \dots, \theta_{m+p-2}^{(j)})$ are independently sampled from a uniform distribution on $[0,1]$ and then perturbed $n_j$ times to produce $n_j$ noisy version. The added noise is sampled from a uniform distribution on $[-\varepsilon, \varepsilon]$, where $\varepsilon > 0$ controls the noise intensity. For the $r$-th perturbation $(r \in \{1, \dots, n\})$, the perturbed coefficients are given by:

\begin{equation*}        
\theta^{(j,r)}_l = \theta_l^{(j)} + \eta_l^{(j,r)} \textrm{, with } \theta_l^{(j)} \sim \mathcal{U}(0,1), \quad \eta_l^{(j,r)} \sim \mathcal{U}(-\varepsilon, \varepsilon)
\end{equation*}

\begin{figure}
    \centering
    \subfloat[Splines simulation, $\varepsilon = 0.2$.]{
      \includegraphics[width=0.45\textwidth]{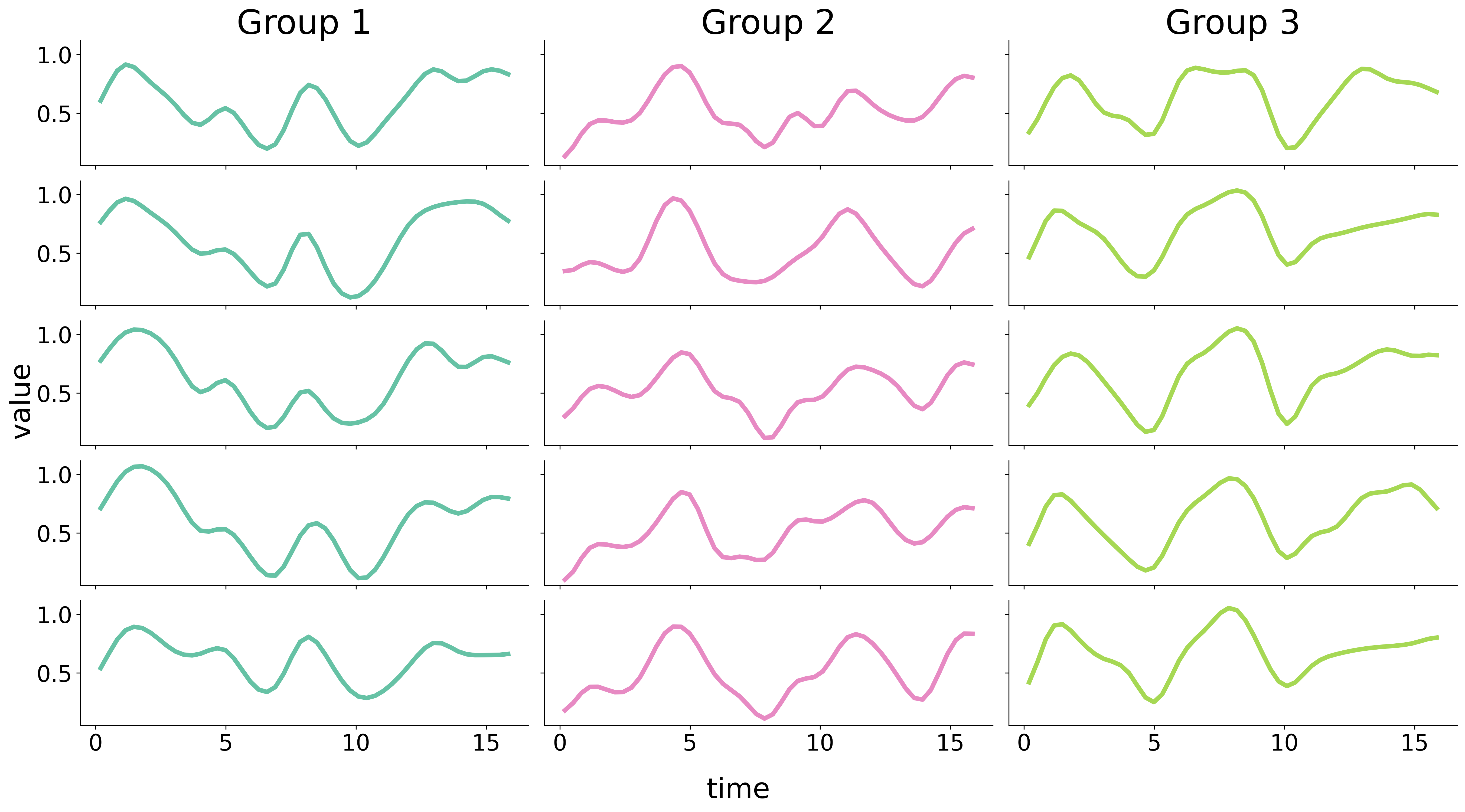}
      \label{splines}
    }
    \hfill
    \subfloat[Histograms simulation, $c = 1000$.]{
      \includegraphics[width=0.45\textwidth]{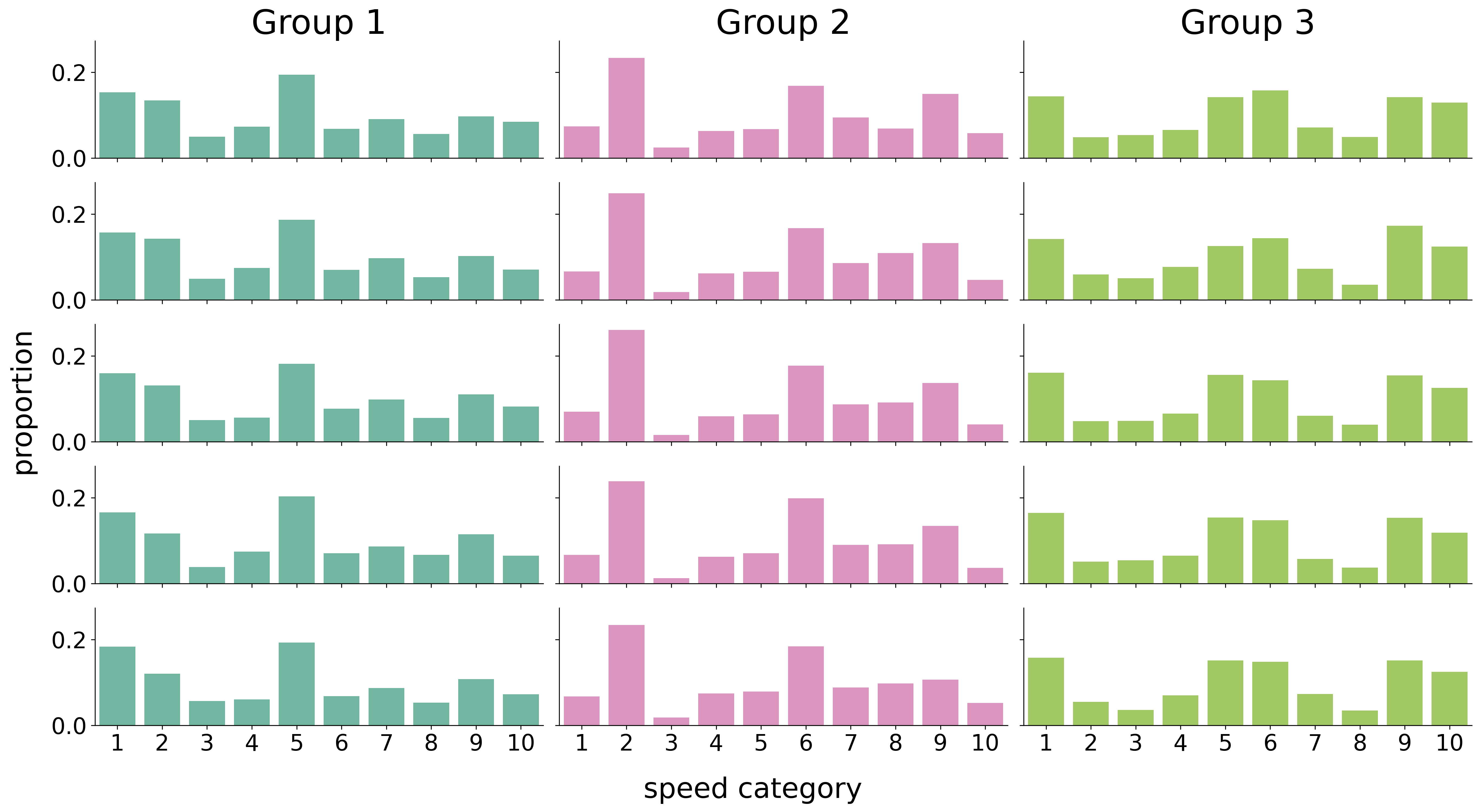}
      \label{histograms}
    }
    \caption{Example of attribute simulation for 3 groups, with 5 individuals per group.}
    \label{fig:example_sim}
\end{figure}

In our simulation settings, each node $i$ (belonging to group $j$) is associated with a functional attribute $f_i$, expressed in the B-spline basis of degree $p=3$ with $m=23$ fixed knots: $f_i(x) = \sum_l^{23+3-2} \theta_l^{j,i} B_{l,3}(x)$. An illustration is given on Figure \ref{splines}.

\paragraph{Histograms Generation}
A Dirichlet distribution $\text{Dir}(\gamma_1, \dots ,\gamma_S)$ with a support size $S$ is considered. The concentration parameters $\gamma_s$ for $s=1,\dots,S$ are drawn independently from a uniform distribution. Sampling once from this Dirichlet yields a probability vector $p^{(1)} = (p^{(1)}_1, \dots, p^{(1)}_S)$,
where each component $p^{(j)}_s$ represents the probability mass assigned to the $s$-th category in this realization. Repeating this procedure $k$ times produces $k$ distinct base distributions 
$p^{(1)}, \dots, p^{(k)}$, each corresponding to a different random configuration of category probabilities. For each of these $k$ base distributions $p^{(j)}$, $n_j$ additional variants are generated. Each variant is sampled from a new Dirichlet distribution $\text{Dir}(p_1^{(j,r)}, \dots ,p_S^{(j,r)})$, where the concentration parameters are defined as:

\begin{equation*}
    p_s^{(j,r)} = c \cdot p_s^{(j)}, \quad \textrm{ for } s=1, \dots, S, \quad c \in \mathbb{R}^+
\end{equation*}

For a general Dirichlet distribution $\text{Dir}(\alpha_1, \dots, \alpha_S)$ with $\alpha_0 = \sum_{i=1}^S \alpha_i$,  the variance of the $i$-th component is $\operatorname{Var}[X_i] = \frac{\alpha_i (\alpha_0 - \alpha_i)}{\alpha_0^2 (\alpha_0 + 1)}$. In our case, since $\sum_{i=1}^S p_i^{(j)} = 1$, one has $\alpha_0 = c$ and $\alpha_i = p_i^{(j,r)} = c \cdot p_i^{(j)}$, so that $\operatorname{Var}[X_i] = \frac{p_i^{(j)} (1 - p_i^{(j)})}{c + 1}$. Thus, $c$ acts as a scaling parameter: larger values of $c$ produce variants closer to the original realization $(p_1^{(j)}, \dots, p_S^{(j)})$, while smaller values introduce greater variability. An illustration is given on Figure \ref{histograms} for $c=1000$.

Finally, for each node $i$ (belonging to group $j$) in the graph, the discrete probability distribution $\mathbf{h_i}$ is associated, sampled from the corresponding Dirichlet variant ($\text{Dir}(p_1^{(j,i)}, \dots ,p_S^{(j,i)})$) associated with that node. 

\paragraph{Attributes Distance}
Traffic data may exhibit slight time shift in their peaks across different geographical zones, but the primary objective is to detect similarity in shape. Therefore, Dynamic Time Warping (or DTW) is employed, denoted $d_{\text{dtw}}(f_i, f_j)$, where $f_i$ and $f_j$ are the functional data associated to nodes $i$ and $j$, respectively. Histograms data are compared using the Wasserstein 1-distance, denoted $W_1(\mathbf{h_i}, \mathbf{h_j})$, where $\mathbf{h_i}$ (resp. $\mathbf{h_j}$) is the discrete distribution associated to node $i$ (resp. node $j$). To enable comparison, the attribute distances are also rescaled using min-max normalization, denoted respectively by $\tilde{d}_{\text{dtw}}(f_i, f_j)$ and $\tilde{W_1}(\mathbf{h_i}, \mathbf{h_j})$. Ultimately, the attributes distance are combined: $d_A(v_i, v_j) = \beta \cdot \tilde{d}_{\text{dtw}}(f_i, f_j) + (1 - \beta) \cdot \tilde{W_1}(\mathbf{h_i}, \mathbf{h_j})$, with $\beta \in [0, 1]$ a weighting parameter, set here to $\frac{1}{2}$.

\paragraph{Attributes Barycenters}

For the srFGW approach, attribute barycenters must be computed to define $B$, the attributes of the target graph, and to compute $\mathbf{M}$, the distance matrix between the attributes of the source and target graphs. In this work, the barycenters are computed among the nodes of the graph, which reduces computational cost. 

\paragraph{Perturbation levels}

\begin{figure}
  \centering
  \subfloat[ Attribute perturbation for functional data.\label{perturbation} ]{
    \includegraphics[width=0.9\linewidth]{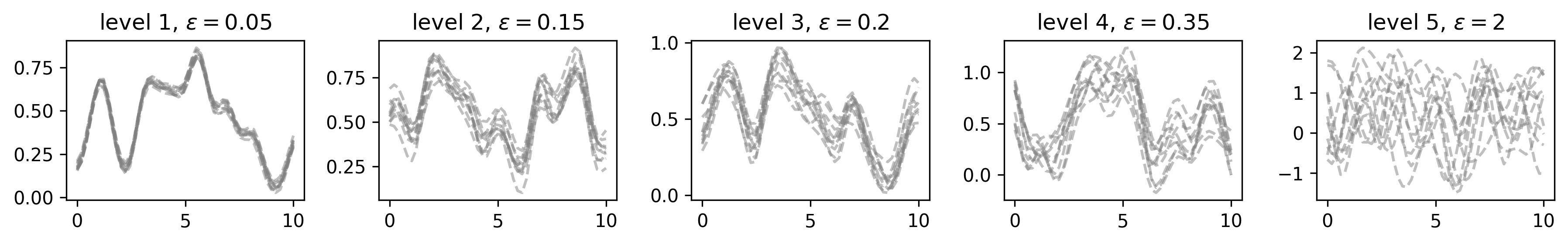}
  }\\[1em]
  \subfloat[ Perturbation-level values.\label{perturbation_levels} ]{
    \begin{tabular}{ccc}
      \hline
      Perturbation level & $\varepsilon$ (function perturbation) & $c$ (histograms perturbation) \\
      \hline
      1 & 0.05 & 1000 \\
      2 & 0.15 & 200 \\
      3 & 0.20 & 80 \\ 
      4 & 0.35 & 15 \\
      5 & 2.00 & 2 \\
      \hline
    \end{tabular}
  }
  \caption{Perturbation levels and their graphical illustration.}
  \label{fig:perturbation-combined}
\end{figure}

Several levels of attribute perturbation have been defined, representing varying degrees of noise among individuals within the same community that share a common underlying distribution or function. These different levels are designed to simulate varying intra-class attribute variances. This principle is illustrated in Figure \ref{perturbation}, and the corresponding values are detailed in Table \ref{perturbation_levels}. The case involving a small perturbation on one attribute type and a large difference on the other is not considered here, as it falls outside the scope of this study.

\subsubsection{Generation of Topology-Aligned Attribute Dissimilarities}

The focus here is on graphs with prescribed topological structures, like chained graphs. To ensure that node attributes follow the same organization, attribute information is generated directly as a dissimilarity matrix which mirrors the underlying graph topology. This construction provides a controlled setting in which the consistency between structural and attribute information can be precisely tuned.

Let $C^{(i)} \in\{1,\ldots,k\}$ denote the group membership of node $i$, and let
$\mathbf{A}$ denote the binary adjacency matrix induced by the SBM block probability matrix, where $a_{rs}=1$ if the corresponding inter-group connection probability is non-zero. The attribute dissimilarity matrix $\mathbf{D}_A$ is constructed by sampling independently each entry $d_A(i,j)$ according to:
\begin{equation*}
d_A(i,j) = d_A(j,i) =
\begin{cases}
0, & \text{if } i=j, \\
\max\{0, Z_{ij}\}, & \text{if } i\neq j,
\end{cases}
\end{equation*}
where $Z_{ij}$ is a random variable sampled as
\begin{equation*}
Z_{ij} \sim
\begin{cases}
\mathcal{N}(\mu_{\mathrm{intra}}, \sigma_{\mathrm{intra}}^2),
& \text{if } C^{(i)}=C^{(j)}, \\
\mathcal{N}\!\left(\dfrac{\mu_{\mathrm{intra}}+0.5}{2},
\left(\dfrac{\sigma^2_{\mathrm{intra}}+0.1}{2}\right)\right),
& \text{if } C^{(i)}\neq C^{(j)}\ \text{and}\ A_{C^{(i)},C^{(j)}}=1,\\
\mathcal{N}(0.5, 0.1),
& \text{otherwise}.
\end{cases}
\end{equation*}

The resulting matrix is finally normalized by its maximum value. This construction yields a dissimilarity matrix whose block structure is aligned with the underlying SBM connectivity: intra-group pairs are the closest, connected inter-group pairs exhibit intermediate distances, and non-connected groups are well separated. Remark that this is not a true distance matrix, since the triangle inequality is not guaranteed, however it is not an issue in this framework. 

\paragraph{Perturbation levels}

As in the transportation-like attribute setting, different levels of attribute perturbation are considered. The corresponding values are reported in Table~\ref{perturbation_levels_topo} and are calibrated based on the mean and standard deviation observed in the perturbation levels previously used for generating transportation-like attributes.

\begin{table}
    \centering
    \begin{tabular}{ccc}
      \hline
      Perturbation level & $\mu_{\mathrm{intra}}$ & $\sigma^2_{\mathrm{intra}}$ \\
      \hline
      1 & 0.1 & 0.02 \\
      2 & 0.2 & 0.04 \\
      3 & 0.25 & 0.06 \\ 
      4 & 0.35 & 0.08 \\
      5 & 0.45 & 0.1 \\
      \hline
    \end{tabular}
  \caption{Perturbation levels for topology-aligned attribute distances.}
  \label{perturbation_levels_topo}
\end{table}

\subsubsection{Evaluation}

Evaluation is performed using ARI. The simulation framework follows the same setup as previously described: each node is assigned to the class receiving the highest mass in the transport plan (breaking ties arbitrarily), and the methods are tested on graphs with 200 nodes distributed across 5 communities of similar size. Monte Carlo simulations with 100 repetitions per setting are conducted, for $\alpha = \frac{1}{2}$, meaning that structure and attributes are equally weighted. Five levels of attribute perturbation (as previously defined) and ten levels of structural strength $\rho$ are explored to assess how the methods perform under varying intra- and inter-community connection ratios.

\subsubsection{Influence of the Source and Target Choices for Fused Gromov--Wasserstein} \label{fgw_section}

\begin{figure}
    \centering
     \begin{subfigure}{\textwidth}
         \centering
         \includegraphics[width=1\textwidth]{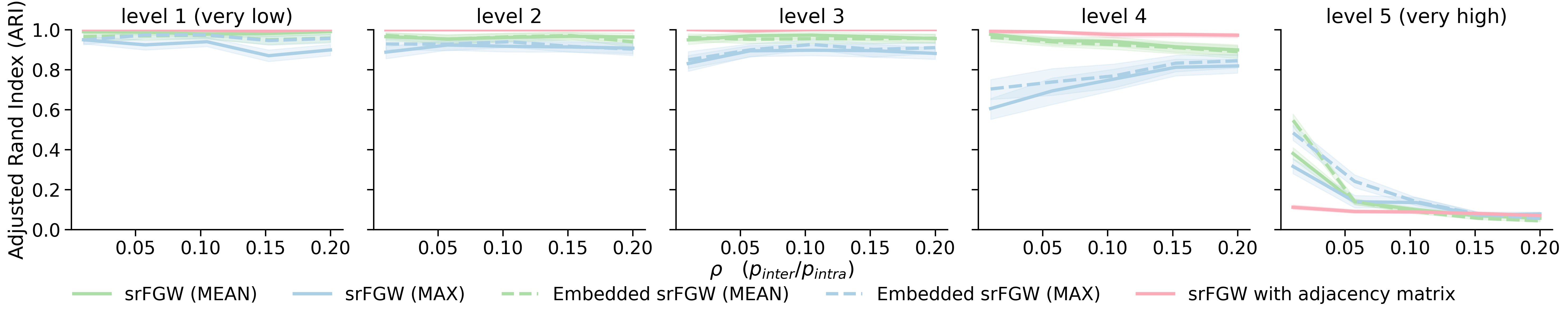}
        \caption{Fully connected graph (synthetic attributes).}
        \label{fused_comparisonA}
     \end{subfigure}

    \vspace{0.5cm}

     \begin{subfigure}{\textwidth}
         \centering
         \includegraphics[width=1\textwidth]{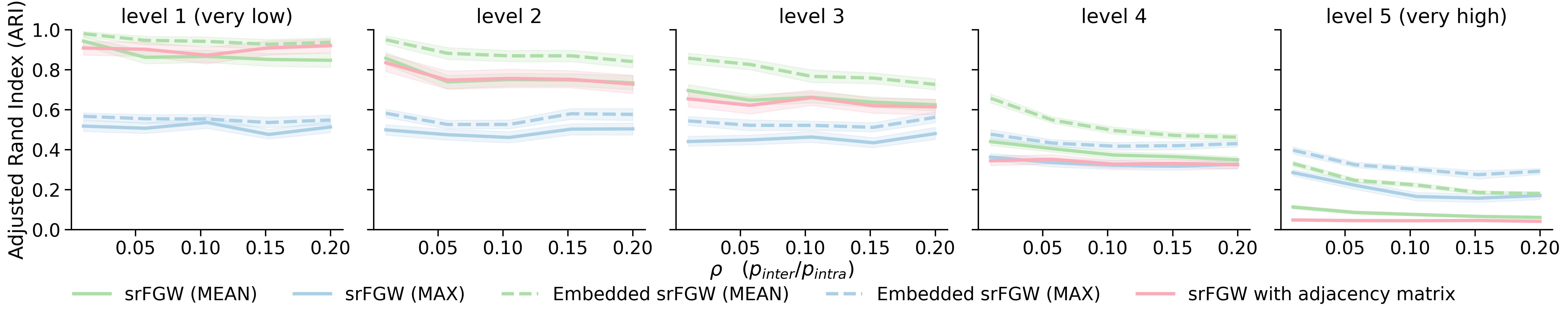}
        \caption{Chain graph (topology-aligned attribute distances).}
        \label{fused_comparisonB}
     \end{subfigure}

    \caption{Comparison of semi-relaxed Fused Gromov--Wasserstein (srFGW) methods across different source representations for the structural component and across different graph topologies.}
    \label{srfgw}
\end{figure}

Like Gromov--Wasserstein, Fused Gromov--Wasserstein raises the question of the choice of source and target representations for its structural component. Figure~\ref{srfgw} compares these choices, in particular when using either the adjacency matrix or the distance matrix as the source representation.

In the fully connected setting, using the adjacency matrix as the source representation outperforms the distance-based alternative. However, in settings with more diverse distance distributions, such as the chain setting, where the adjacency matrix does not fully capture relations between nodes, both adjacency- and non-embedded distance-based strategies achieve similar performance. Moreover, the distance-based formulation exhibits greater robustness to attribute perturbations. Adjacency-based srFGW appears to be strongly affected by attribute perturbations and performs poorly when attribute distances contain little informative signal (at the highest perturbation level), both in fully connected settings and in other graph topologies.

For the distance-based strategy, using either the mean or the maximum source distance to define inter-node distances in the target graph leads to similar conclusions as in the previous section: the mean-based choice generally yields better performance. However, the maximum-based alternative, despite its weaker overall accuracy, appears to be more robust to attribute noise.

Using the embedded representation generally yields comparable or improved performance in srFGW, with the largest gains observed in challenging settings where attributes provide limited information. In complex topologies, such as the chain setting (Figure~\ref{fused_comparisonB}), the improvement brought by the embedding is particularly pronounced, and the embedded variant with a mean-distance target consistently outperforms the adjacency-based srFGW formulation. Table~\ref{attributed_kernel_comparison} further shows that this improvement is more pronounced in the fused formulation than in srGW, where, as highlighted above, results are more contrasted.

\begin{table}
    \centering
    \begin{tabular}{lcccc}
    \hline
     & \multicolumn{2}{c}{Level 1 (very low)} & \multicolumn{2}{c}{Level 5 (very high)} \\
    \cline{2-3} \cline{4-5}
     & - & embedded & - & embedded \\
    \hline
    srGW (mean)      & \textbf{1.00} & 0.986 & \textbf{0.887} & 0.826 \\
    srGW (max)       & 0.832 & \textbf{0.922} & \textbf{0.623} & 0.574 \\
    \hline
    srFGW (mean)      & 0.865 & \textbf{0.941} & 0.075 & \textbf{0.223} \\
    srFGW (max)       & 0.535 & \textbf{0.552} & 0.165 & \textbf{0.301} \\
    \hline
    \end{tabular}
    \caption{Comparison (average ARI) of embedded and non-embedded sr(Fused)GW (chain graph, $\rho=0.1$).}
    \label{attributed_kernel_comparison}
\end{table}

\subsubsection{Comparison of Gromov--Wassertein Methods}

\begin{figure}
    \centering
    \begin{subfigure}{1\textwidth}
        \centering
        \includegraphics[width=1\textwidth]{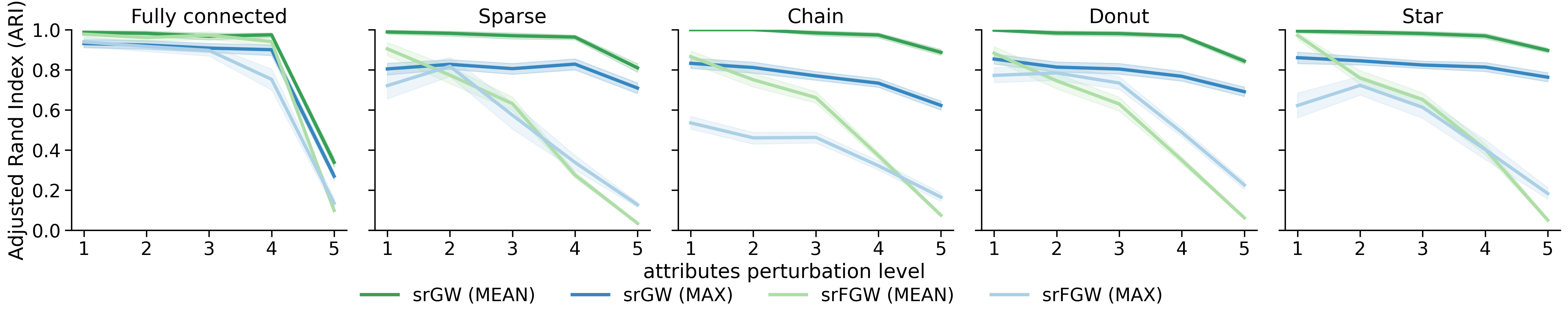}
        \caption{Comparison of semi-relaxed Fused and semi-relaxed Gromov--Wasserstein partitioning.}
        \label{gw_fused_comparison}
    \end{subfigure}
    
    \vspace{0.5cm}
    
    \begin{subfigure}{1\textwidth}
        \centering
        \includegraphics[width=1\textwidth]{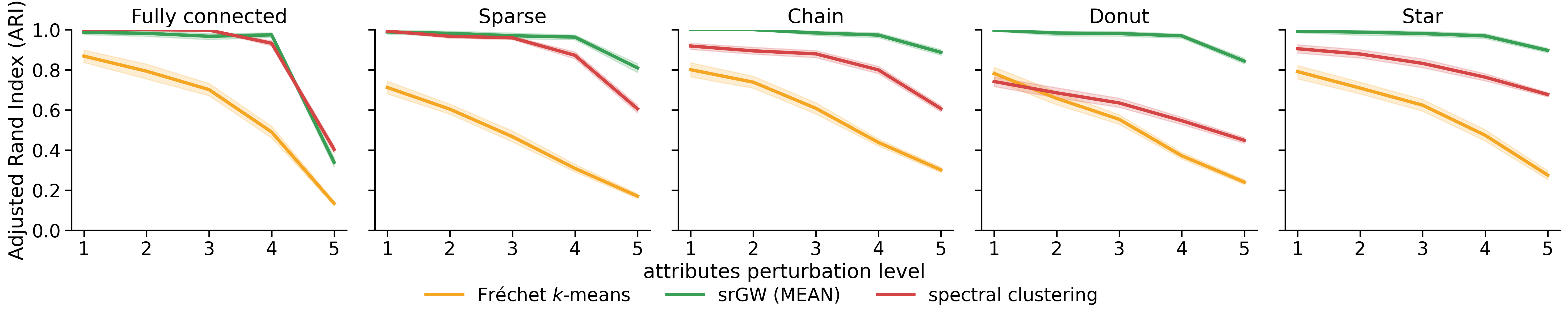}
        \caption{Comparison of the best-performing semi-relaxed Gromov--Wasserstein method with non-Gromov-based methods.}
        \label{other_attributed_comparison}
    \end{subfigure}
    \caption{Comparison of partitioning performance (average ARI) of methods across varying levels of attributes perturbation and various shapes ($\rho=0.1$).}
\end{figure}

Figure~\ref{gw_fused_comparison} highlights that in this setting, srFGW (in light colors) achieves performance comparable to srGW in the fully connected case, but is strongly degraded for other graph topologies. As attribute perturbations increase, this observation becomes even more pronounced, whereas srGW methods appear to be robust to attribute noise. This raises the question of the choice of $\alpha$, the weighting parameter between structural and attribute information, which is fixed to $0.5$ in this study but could benefit from being adapted.

\begin{figure}[htbp]
    \centering
    \begin{minipage}{0.6\textwidth}
        \centering
        \includegraphics[width=\linewidth]{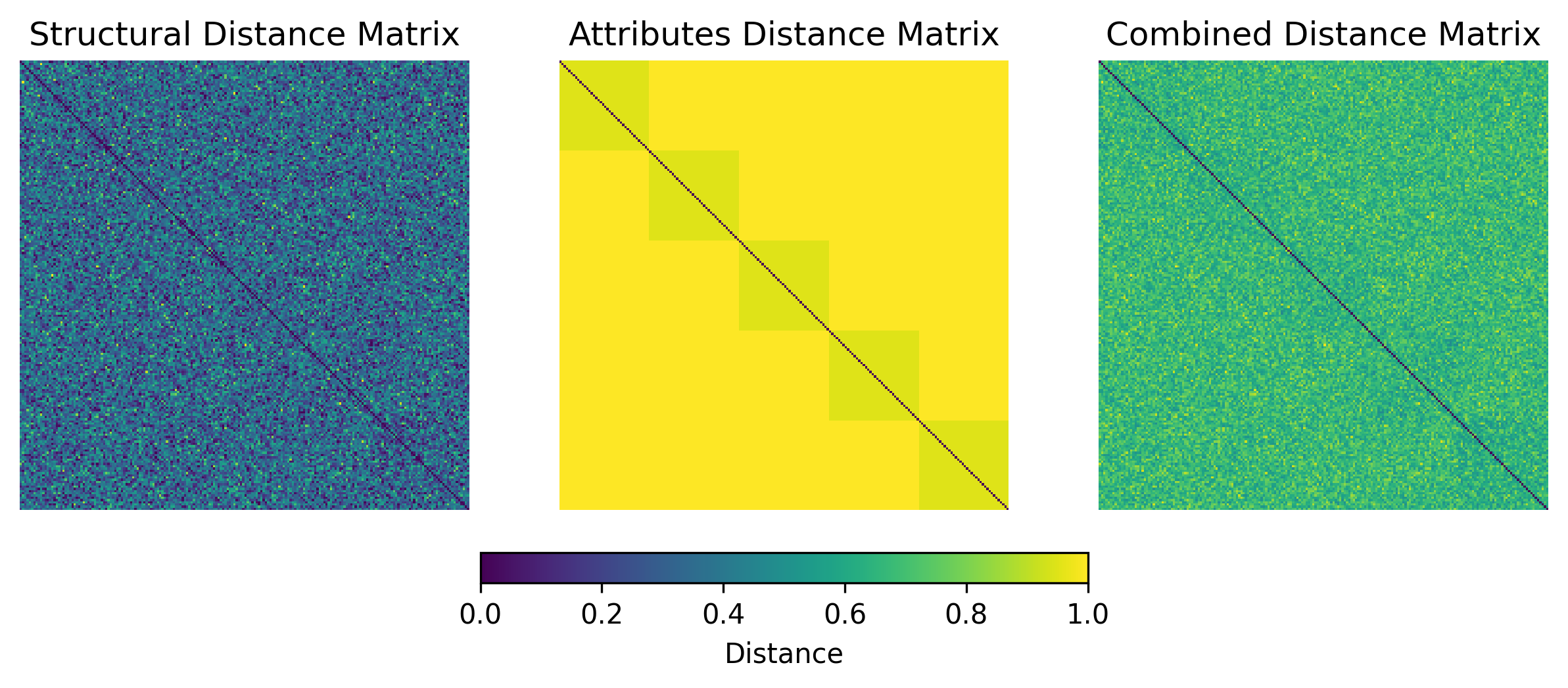}
    \end{minipage}
    \hfill
    \begin{minipage}{0.35\textwidth}
        \centering
        \begin{tabular}{lr}
            \hline
             & ARI (average) \\
            \hline
            srGW (mean)  & 0.434 \\
            srFGW (mean) & \textbf{0.829} \\
            \hline
        \end{tabular}
    \end{minipage}
    \caption{Comparison of performance (average ARI) in a setting with no structural information and only attribute structure.}
    \label{fused_performance}
\end{figure}

Figure~\ref{fused_performance} illustrates an example that is not captured by our simulation framework: there is no structural information at all (completely noisy structural distance matrix), but some attribute information is present. Using this combined distance matrix in this experimental setting, srFGW performs substantially better than srGW, which could be attributed to srFGW optimization strategy, optimizing attributes and structure separately.

\subsubsection{Comparison with Non-Gromov-Wasserstein Method}

As in the non-attributed setting, the best-performing Gromov--Wasserstein-based method is compared with Fr\'{e}chet $k$-means and spectral clustering. While Fr\'{e}chet $k$-means can operate directly on the combined distance matrix $\mathbf{D}_{\alpha}$, spectral clustering requires a similarity matrix. Therefore, a similarity matrix $\mathbf{S}_{\alpha}$ is constructed from $\mathbf{D}_{\alpha}$ using a Gaussian kernel \parencite{ng2001spectral}:

\begin{equation*}
  (\mathbf{S}_{\alpha} )_{ij}
    =
    \exp(-\frac{d_\alpha(i,j)^{\,2}}{2\sigma^2}), \quad \text{with} \qquad
    \sigma = \mathrm{median}(\mathbf{D}_\alpha).
\end{equation*}

Results are presented in Figure~\ref{other_attributed_comparison} for a moderate intra-/inter-group connectivity ratio ($\rho = 0.1$). Fr\'{e}chet $k$-means yields poorer performance than the other two methods, and this alternative distance-based method is strongly affected by attribute perturbations. 

On fully connected graphs, spectral clustering and srGW achieve similar results and exhibit comparable behaviour with respect to attribute perturbations. In contrast, spectral clustering appears to be more sensitive to changes in graph topology, with a marked performance degradation, particularly in the donut setting, whereas srGW maintains a similar level of performance across the different topologies. Moreover, for srGW, attribute perturbations only begin to affect performance when the attributes become nearly random and therefore carry almost no discriminative information. Otherwise, the method remains robust and its performance is mostly unaffected.

%% file: sections/4_applications.tex
\FloatBarrier
\section{Application to Real-World Traffic Data}

In addition to simulated data, we apply the proposed graph partitioning methods to real-world data. The dataset comes from traffic observations and includes, in particular, speed distributions and weekly traffic flow curves for streets, derived from Floating Car Data. Floating Car Data (FCD) refers to information collected from vehicles equipped with GPS or other tracking technologies.

\subsection{Road Network Data Description}

The FCD used in this study pertains to the city of Châteaubourg (Brittany, France), a small town of approximately 7,500 residents that experiences recurrent congestion issues. This city was selected due to an ongoing citizen science project, that aims to measure traffic to better understand car mobility in low-density areas and support informed public policy decisions.

\begin{figure}
        \centering
        \subfloat[Cul-de-sac -- Primal.]{
            \includegraphics[width=0.2\linewidth]{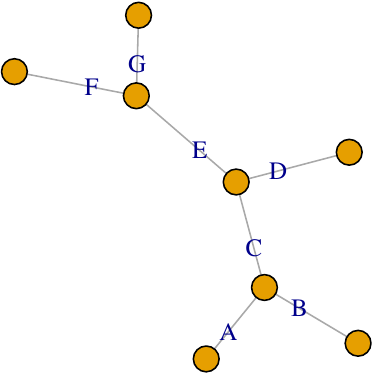}
            \label{fig:culdesac_primal}
        }
        \hfill
        \subfloat[Cul-de-sac -- Dual.]{
            \includegraphics[width=0.2\linewidth]{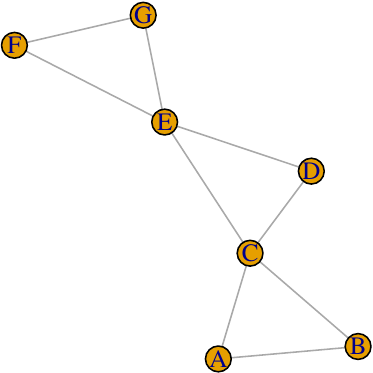}
            \label{fig:culdesac_dual}
        }
        \hfill
        \subfloat[Roundabout -- Primal.]{
            \includegraphics[width=0.2\linewidth]{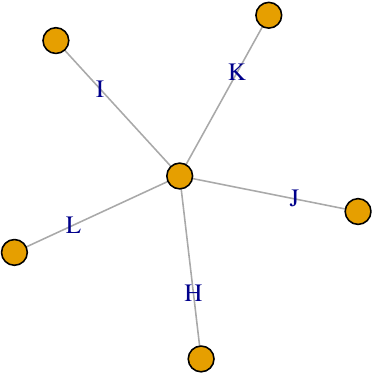}
            \label{fig:roundabout_primal}
        }
        \hfill
        \subfloat[Roundabout -- Dual.]{
            \includegraphics[width=0.2\linewidth]{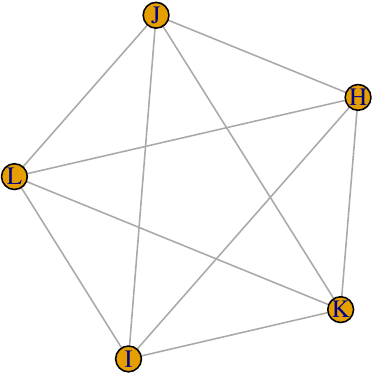}
            \label{fig:roundabout_dual}
        }
        \caption{Illustration of dual network.}
        \label{dual_illustration}
\end{figure}

\begin{minipage}{0.6\textwidth}
\paragraph{Dual network} The road network of Châteaubourg (Figure \ref{ctb_network}) is represented in our framework by a graph composed of 831 streets segments (corresponding to portions of streets), connected through 748 intersections. These intersections correspond not only to physical junctions, such as roundabouts or crossroads, but also to simple connections between consecutive streets portions. In this representation, intersections serve as nodes and street segments as edges. However, since the goal is to partition the streets rather than the intersections, the dual graph representation is adopted, which is widely used in the transportation literature. \parencite{lin2013complex}. In the dual graph, each road segment is represented as a node, and an undirected edge is established between two nodes whenever their corresponding street segments share at least one intersection in the original network. These edges thus encode the adjacency between road segments.  Figure \ref{dual_illustration} illustrates this transformation from the primal graph of example networks (left) to their corresponding dual graph (right). Several changes may be observed, notably an increased number of triangles.
\end{minipage}%
\hfill
\begin{minipage}{0.35\textwidth}
    \centering
    \includegraphics[width=0.8\linewidth]{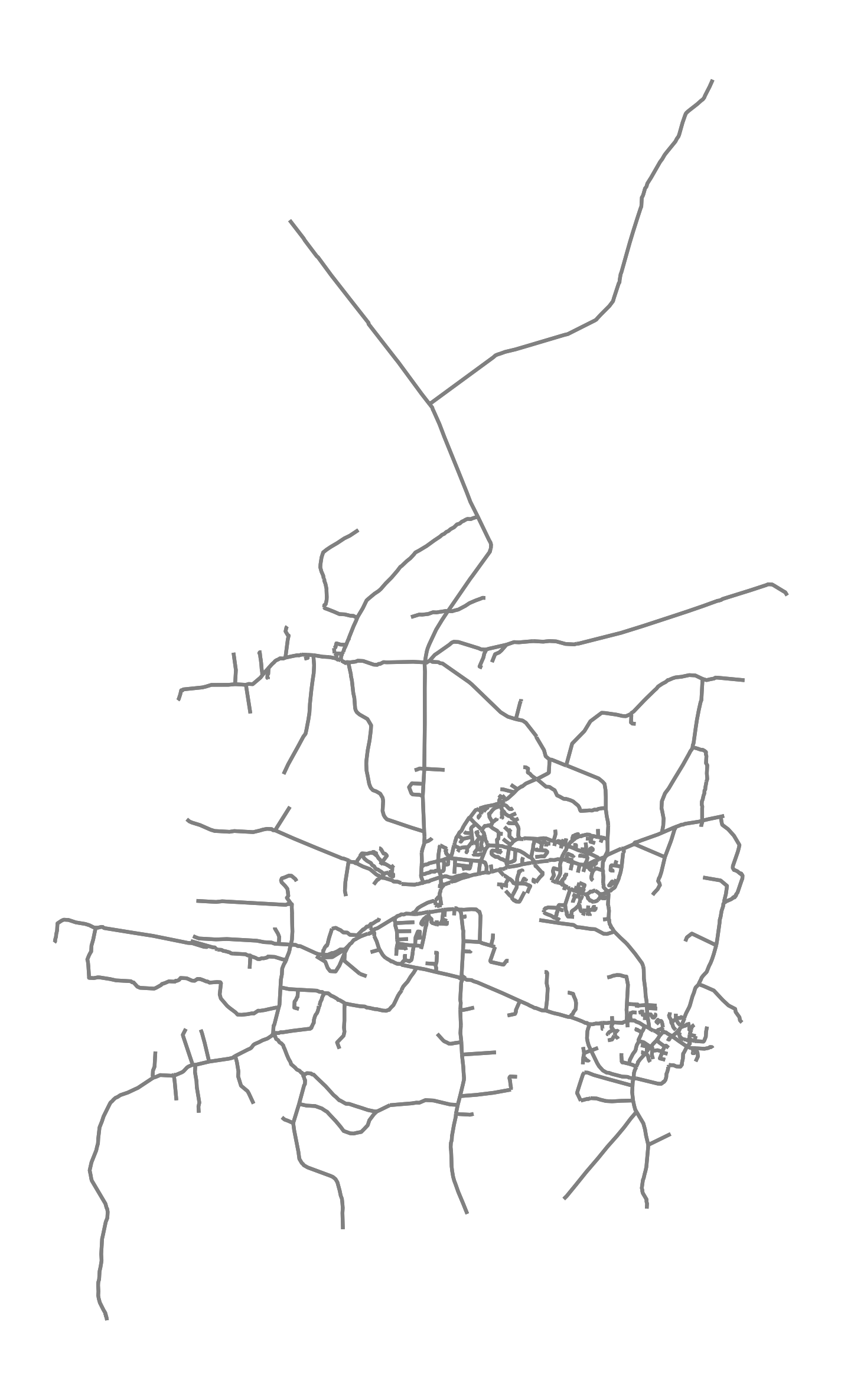}
    \captionof{figure}{Road network of Châteaubourg}
    \label{ctb_network}
\end{minipage}

As in the primal representation, the dual construction does not preserve node coordinates within a geographic reference system. Moreover, since edges in the dual graph represent intersections between street segments, they do not inherently carry a notion of length. To retain topological distance information, a weight equal to the mean distance of its two endpoints (i.e., the two street segments it connects) is assigned to each edge of the dual graph. This ensures that a relevant notion of distance is preserved across the dual representation.

\subsubsection{Data Details} Raw FCD data collected from 2021 to 2024 provide speed statistics and vehicle counts. For this study, we use a pre-processed and resampled dataset with traffic curves (number of vehicles per 15-minutes interval), and speed histograms (with 5 km/hour bins), both aggregated by road segments. While histograms are computed over the entire observation period, traffic curves are restricted to the year 2022, which offers the highest data quality, and further aggregated to form a representative weekly curve for the whole year. Traffic counts are normalized by road section and direction to prevent the analysis from only reflecting overall traffic volume differences and to enable a more fine-grained interpretation. Finally, although data are available in both traffic directions, a modeling choice is adopted in which each road segment is represented by a single node, with attributes combining information from both directions. This choice is motivated by the ultimate objective of partitioning streets, while also simplifying the graph construction.

\subsubsection{Attribute Distances} Depending on whether a street is one-way or two-way, each node in the graph is characterized by one or two speed histograms and one or two weekly traffic curves. Consequently, the attribute distances used in our simulations must be adapted to compare subsets of attributes. The Hausdorff distance is specifically designed to measure distance between two subsets $A$ and $B$, and numerous variations have been introduced with the purpose of object matching \parencite{dubuisson1994modified}. The original Hausdorff distance is a true metric on the set of nonempty closed bounded subsets of a general metric space, but some of its variants are not. Generally, this would not be an issue for our application. In this study, the following variant is adopted:
\begin{equation*}
    d(a, B) = \min_{b \in B} d_\textrm{A}(a,b), \quad 
    \textrm{HD}(A, B) = \frac{1}{2} \Big( \max_{a \in A} d(a, B) + \max_{b \in B} d(b, A) \Big)
\end{equation*}

With this variant, both directions of a road are considered equally, while differences between the subsets are penalized. As in the simulations, the attribute distance $d_\textrm{A}(\cdot,\cdot)$ is computed using the DTW distance for traffic curves and the Wasserstein-1 distance for speed histograms. Since the variances of length, traffic, and speed distances are large, a square root transformation is applied to each distance matrix in this application. This adjustment helps to prevent trivial clusterings (e.g., simply separating low-speed from high-speed roads) and allows for a more balanced combination of the different distance types. Note that the transformed matrices are not guaranteed to satisfy all the properties of distance matrices; however, this isn't an issue for our methods.

\subsection{Results}

\subsubsection{Discussion of Methods Settings}

\paragraph{Initialization}

Initialization plays a crucial role in all distance-based methods presented in this paper and compared in this section. The methods are initialized using a $k$-means++ procedure applied to the embedded matrix, repeated 100 times. The results presented correspond to the initialization that minimizes the objective function of each method.

\paragraph{Number of Clusters}

Choosing the number of clusters is a widely studied problem in the literature \parencite{mirkin2011choosing}, and various strategies have been proposed, such as the silhouette index, which balances the within-cluster and between-cluster distances, or variance-based approaches. In our specific application, the goal is to distinguish zones with characteristic traffic patterns, which suggests a limited number of clusters. Using the same number of clusters across methods allows for a clearer illustration of their differences on real data, although it does not facilitate selecting the optimal number of clusters according to a criterion, since such criteria may differ between methods. Therefore, we choose to illustrate the methods using 8 clusters, providing sufficient flexibility to highlight their behavior in this small city.

\paragraph{Evolution with \texorpdfstring{$\alpha$}{alpha}}

\begin{figure}[H]
\centering
\includegraphics[width=0.9\textwidth]{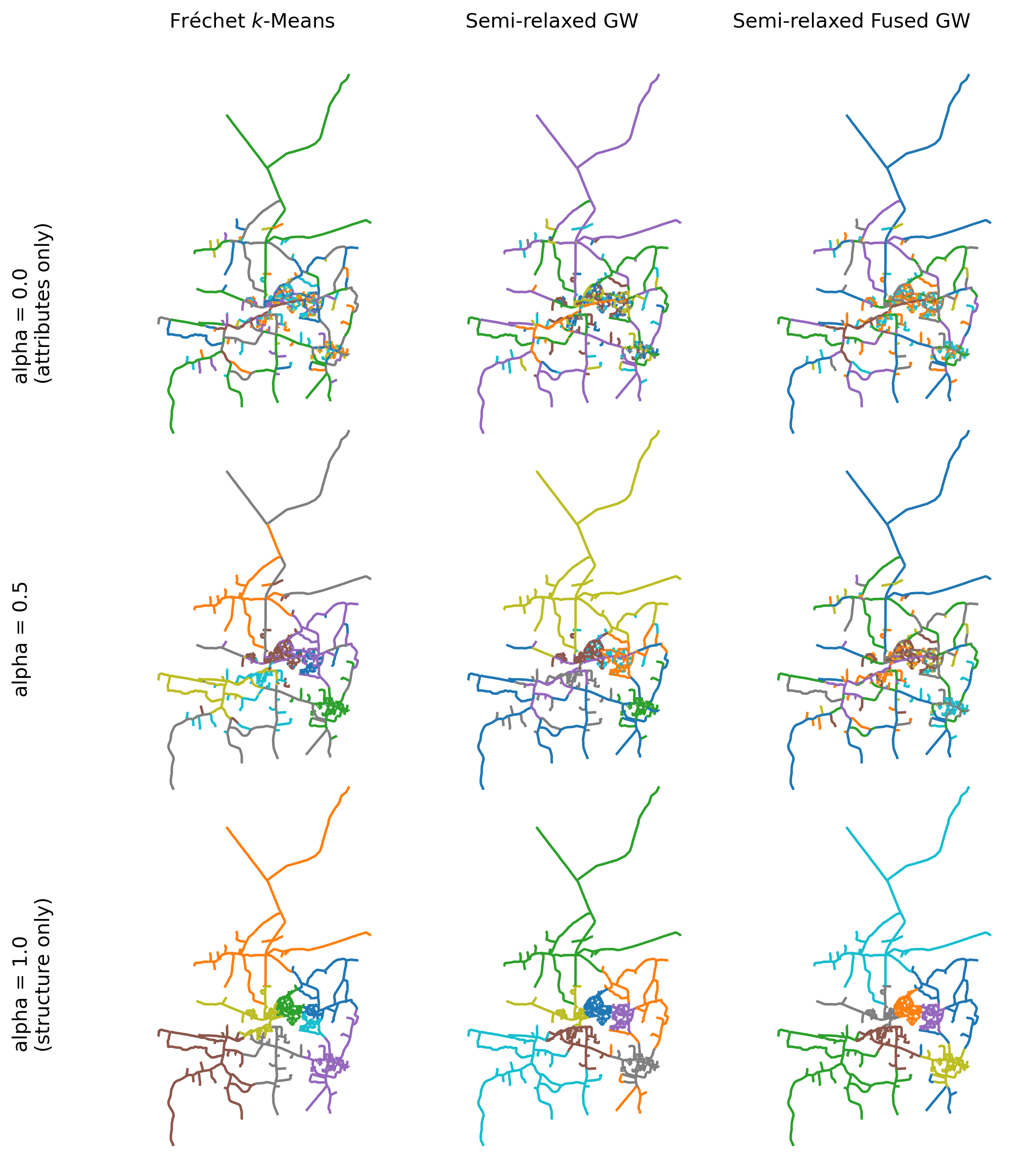}
\caption{Clustering results obtained with Fr\'{e}chet-$k$ means, semi-relaxed Gromov--Wasserstein, and semi-relaxed Fused Gromov--Wasserstein for different values of $\alpha$.}
\label{alpha_results}
\end{figure}

As a reminder, $\alpha$ is the weighting parameter that balances structural and attribute information; for instance, $\alpha = 0$ corresponds to considering only attribute distances. Figure~\ref{alpha_results} presents the clustering results of the different methods, all initialized using the same set of embedded $k$-means++ cluster centers, for the attribute-only, structure-only, and balanced ($\alpha = 0.5$) settings. Clusters are represented by different colors. Note that none of the methods inherently ensures connected subnetworks within the resulting partitions, which is not an issue for our application. However, in applications where connected partitions are desired, the general framework could be further adapted, by introducing, for instance, a regularization term penalizing spatially non-compact clusters, although this aspect is beyond the scope of this present work.

When only the graph structure is considered, this figure illustrates that srFGW reproduces the results obtained with srGW. In this structure-only case, an interesting phenomenon is observed: the resulting subgraphs are not necessarily connected. In particular, peripheral road segments are often grouped together (orange cluster for srGW and blue cluster for srFGW), even when they are not spatially close. This grouping arises because these road segments share similar distance patterns with respect to the rest of the network, rather than direct physical proximity. In other experimental settings, we also observed that peripheral roads can be clustered together even when they are geographically far apart, further illustrating that the method captures similarity in global structural roles rather than local connectivity alone.

\subsubsection{Interpretation in Balanced Settings (\texorpdfstring{$\alpha = 0.5$}{alpha=0.5})} With balancing structural and attribute information, the Fr\'{e}chet-$k$ means and srGW methods mainly partition the city center, while suburban areas are grounded into cluster according to their geographic location. srFGW, however, partitions both the dense city-center streets and suburban road segments based on their traffic behavior. As a result, road sections that are very distant from each other may end up in the same cluster, such in the blue cluster, which corresponds to the main exit from the city, or the green cluster, which gathers secondary roads.

\begin{figure}
\centering
\includegraphics[width=0.8\textwidth]{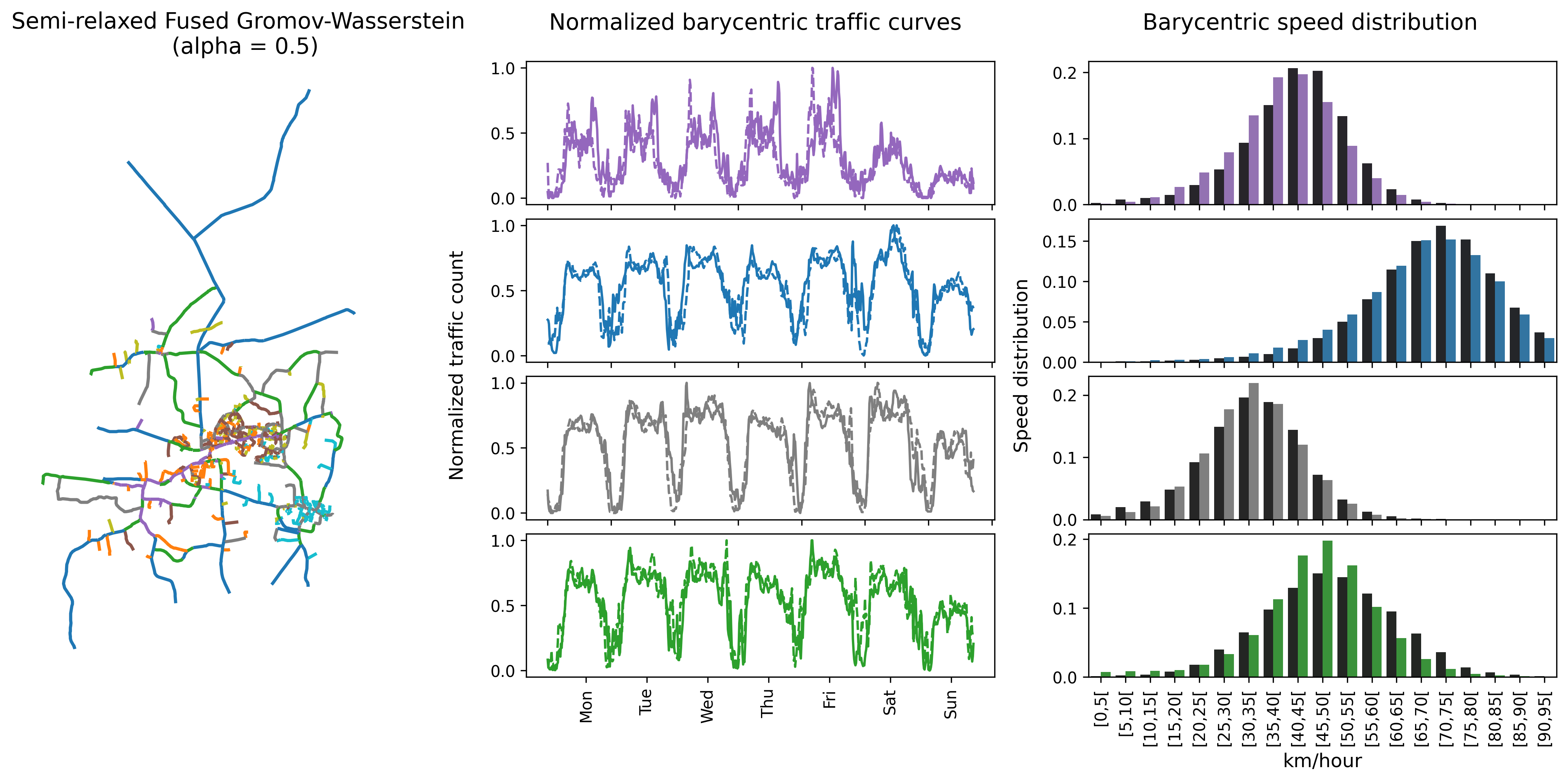}
\caption{Traffic curves and speed distributions of the barycenters of clusters obtained using the semi-relaxed Fused Gromov--Wasserstein method, with a balanced setting ($\alpha = 0.5$).}
\label{srfgw_results}
\end{figure}

Figure~\ref{srfgw_results} provides a detailed view of a selection of clusters obtained in this setting. The barycenter of each cluster is shown for both traffic curves and speed distributions, in both directions. The main road in the city center, colored in purple, exhibits a medium-speed distribution and traffic that decreases on weekends, indicating heavier usage on working days. The speed distribution also shows a slight difference between the two directions of travel, with one direction experiencing more congestion than the other. 

The other three clusters shown in Figure~\ref{srfgw_results} correspond to suburban areas with distinct speed distributions: the blue cluster, representing the main exits from the city, exhibits higher speeds than the grey and green clusters. The latter two display almost identical traffic curves but show a notable difference in their speed distributions. From a traffic perspective, this partition therefore appears to be meaningful, as each group differs from the others in at least one of the following aspects: structural characteristics, traffic or speed patterns.

\FloatBarrier

%% file: sections/X_appendix.tex
\appendix
\section{Proofs of theoretical results}
\label{app:proofs}

\subsection{Proof of Proposition~\ref{prop:loss_decreasing}}
\label{app:proof_loss_decreasing}

\begin{proposition*}
Let $\mathcal{L}(T, B)$ denote the semi-relaxed Fused Gromov-Wasserstein loss 
for a transportation plan $T$ and barycentric attributes $B$. 
If $(T^n,B^n)$ are the iterates produced by Algorithm 3 of the main article, 
then the sequence $(\mathcal{L}(T^n,B^n))_{n\geq 0}$ is monotonically non-increasing. 
More precisely, at each iteration $n$ we have
\begin{equation*}
    \mathcal{L}(T^{n+1}, B^{n+1}) \leq \mathcal{L}(T^{n+1}, B^n) \leq \mathcal{L}(T^n, B^n).
\end{equation*}
\end{proposition*}

\begin{proof}
Let $B = (b_1, \dots, b_k)$ denote the barycentric attributes of a known transportation plan, 
and let $M(B) = [d_A(v_i, b_l)]_{\substack{1 \leq i \leq n \\ 1 \leq l \leq k}}$ be the $n \times k$ distance matrix 
between the attributes of the source graph and $B$. 
Recall that the semi-relaxed Fused Gromov-Wasserstein loss is defined as:
\begin{equation}\label{loss}
    \mathcal{L}(T,B) = \sum_{i,j,l,m} ( (1 - \alpha) d_A(v_i^{(1)}, b_l^{(2)})^q + \alpha |R^{(1)}_{ij} - R^{(2)}_{lm}|^q ) T_{il} T_{jm}.
\end{equation}

At iteration $n$, the algorithm produces iterates $(T^n,B^n)$.  

\textbf{Step 1: Update of the transportation plan.}
For fixed barycenters $B^n$, the algorithm computes 
$T^{n+1} = \arg\min_{T \mathbf{1}_k = \mu} \mathcal{L}(T,B^n)$. 
By optimality of this update, we directly obtain:
\begin{equation}
\label{step1}
  \mathcal{L}(T^{n+1}, B^n) \leq \mathcal{L}(T^n, B^n).
\end{equation}

\textbf{Step 2: Update of the barycentric attributes.}

Equation \ref{loss} can be decomposed into two terms: a attribute-dependent term involving $B$, and a structural term $S_\alpha(T^{n+1})$ independent of $B$.
\begin{align*}
    \mathcal{L}(T^{n+1}, B) &= (1 - \alpha) \sum_l \sum_i T_{il}^{n+1} d_A(v_i, b_l)^q + \alpha \sum_{i,j,l,m} |R_{ij}^{(1)} - R_{lm}^{(2)}|^q T_{il}^{n+1} T_{jm}^{n+1} \\
    &= (1 - \alpha) \sum_l \sum_i T_{il}^{n+1} d_A(v_i, b_l)^q + S_\alpha(T^{n+1})
\end{align*}

The attribute-dependent term is separately minimized with respect to each $b_l$ while keeping $T^{n+1}$ fixed. Each barycenter $b_l^{n+1}$ is chosen to minimize $\sum_{i=1}^n T_{il}^{n+1} d_A(v_i, b_l)^q$. Hence,

\begin{equation}\label{step2}
    \mathcal{L}(T^{n+1},B^{n+1}) \leq \mathcal{L}(T^{n+1}, B^n)
\end{equation}

Combining inequalities \eqref{step1} and \eqref{step2}, we obtain:

\begin{equation*}
     \mathcal{L}(T^{n+1},B^{n+1}) \leq \mathcal{L}(T^{n+1}, B^n) \leq \mathcal{L}(T^n, B^n).
\end{equation*}

Hence, the sequence $(\mathcal{L}(T^n,B^n))_{n\geq 0}$ is monotonically non-increasing.
\end{proof}

\subsection{Proof of Proposition \ref{prop:hard_clustering}}\label{app:hard_clustering}

\begin{proposition*}
Let $(T, B)$ be a solution obtained by Algorithm 3 of the main article. Let $\tilde T$ be another transport plan obtained from $T$ by any projection or modification. For each cluster $C_l:=\{i:\ \tilde T_{il}>0\}$, let the associated barycenter be recomputed as
\begin{equation*}
    \tilde b_l \in \argmin_b \sum_{i\in C_l} \mu_i d_A(v_i,b)^q,
\end{equation*}

Define
\begin{equation*}
D_A:=\max_{\substack{1\le i\le n \\ 1\le l\le k}} d_A(v_i,b_l),
\qquad
D_S:=\max_{\substack{1\le i,j\le n \\ 1\le m,l\le k}}|R^{(1)}_{ij}-R^{(2)}_{lm}|,
\end{equation*}

with respect to the original barycenters $B$.

Then the loss increase induced by replacing $(T,B)$ with $(\tilde T,\tilde B)$ satisfies

\begin{equation*}
\mathcal{L}(\tilde T, \tilde B)-\mathcal{L}(T,B)
 \le
((1-\alpha)D_A^q+2\alpha D_S^{\,q})\ \sum_i \sum_l |T_{il}-\tilde T_{il}|.
\end{equation*}
\end{proposition*}

\begin{proof}

The loss can be written as the sum of an attribute term and a structural term
\begin{equation*}
   \mathcal{L}(T,B)
=(1-\alpha)\sum_{i,l} d_A(v_i,b_l)^q\,T_{il}
+\alpha\sum_{i,j,l,m} c_{ijlm}\,T_{il}T_{jm},
\quad c_{ijlm}:=|R^{(1)}_{ij}-R^{(2)}_{lm}|^q.
\end{equation*}

\textbf{Attribute part.}  
By optimality of the hard barycenters, we have
\begin{equation*}
    \mathcal{L}(\tilde T,\tilde B) \le \mathcal{L}(\tilde T,B)
\end{equation*}

Hence
\begin{equation*}
\mathcal{L}(\tilde T,\tilde B)-\mathcal{L}(T,B)
\ \le\
\mathcal{L}(\tilde T,B)-\mathcal{L}(T,B).
\end{equation*}

For fixed $B$, the attribute contribution is linear in $T$. Hence, the difference between the soft and hard cases is bounded by the largest possible attribute distance:
\begin{equation*}
|(1-\alpha)\sum_{i,l} d_A(v_i,b_l)^q\,(T_{il}-\tilde T_{il})|
\le (1-\alpha)D_A^q \sum_i \sum_l |T_{il}-\tilde T_{il}|,
\end{equation*}

\textbf{Structural part.}  
This term is quadratic in $T$. Consider
\begin{equation*}
\Delta_S = \alpha
\sum_{i,j,l,m} c_{ijlm}\,(
T_{il}T_{jm}
- \tilde T_{il}\tilde T_{jm}).
\end{equation*}

We expand the difference as
\begin{equation*}
T_{il}T_{jm} - \tilde T_{il}\tilde T_{jm}
= (T_{il}-\tilde T_{il})T_{jm}
+ \tilde T_{il}(T_{jm}-\tilde T_{jm}).
\end{equation*}

Taking absolute values and using the triangle inequality yields
\begin{equation*}
|T_{il}T_{jm} - \tilde T_{il}\tilde T_{jm}|
\le |T_{il}-\tilde T_{il}|\cdot T_{jm}
+ |T_{jm}-\tilde T_{jm}|\cdot \tilde T_{il}.
\end{equation*}

Since both $T$ and $\tilde T$ are transport plans with total mass equal to 1. Hence, we obtain the general bound 
\begin{equation*}
  \sum_{i,j,l,m}
|T_{il}T_{jm} - \tilde T_{il}\tilde T_{jm}|
 \le
2 \sum_{i,l}|T_{il}-\tilde T_{il}|.
\end{equation*}

Finally, multiplying by $\max c_{ijlm} = D_S^q$ and by $\alpha$ gives
\begin{equation*}
|\Delta_S| \le 2\alpha D_S^q
\sum_{i,l} |T_{il}-\tilde T_{il}|.
\end{equation*}

\textbf{Combination.}  
Combining the attribute and structural bounds shows that
\begin{equation*}
\mathcal{L}(\tilde T,\tilde B)-\mathcal{L}(T,B) \le
((1-\alpha)D_A^q+2\alpha D_S^q)
\sum_{i=1}^n \sum_{l=1}^k |T_{il}-\tilde T_{il}|.
\end{equation*}
\end{proof}

